\documentclass[preprint2]{aastex6}

\usepackage{amssymb}
\usepackage{soul}
\usepackage{graphicx}
\usepackage{amsmath}
\begin{document}
\sloppy

\title{Solar System objects observed with TESS -- First data release: bright main-belt and Trojan asteroids from the Southern Survey}
\shorttitle{TSSYS-DR1: Solar System objects as observed by TESS: first data release}

\author{Andr\'as P\'al\altaffilmark{1,2,3}}
\author{R\'obert Szak\'ats\altaffilmark{1}}
\author{Csaba Kiss\altaffilmark{1}}
\author{Attila B\'odi\altaffilmark{1,4}}
\author{Zs\'ofia Bogn\'ar\altaffilmark{1,4}}
\author{Csilla Kalup\altaffilmark{2,1}}
\author{L\'aszl\'o L. Kiss\altaffilmark{1}}
\author{G\'abor Marton\altaffilmark{1}}
\author{L\'aszl\'o Moln\'ar\altaffilmark{1,4}}
\author{Emese Plachy\altaffilmark{1,4}}
\author{Kriszti\'an S\'arneczky\altaffilmark{1}}
\author{Gyula M. Szab\'o\altaffilmark{5,6}}
\author{R\'obert Szab\'o\altaffilmark{1,4}}

\email{apal@szofi.net}
\altaffiltext{1}{Konkoly Observatory, Research Centre for Astronomy and Earth Sciences, H-1121 Budapest, Konkoly Thege Mikl\'os \'ut 15-17, Hungary}
\altaffiltext{2}{E\"otv\"os Lor\'and University, H-1117 P\'azm\'any P\'eter s\'et\'any 1/A, Budapest, Hungary}
\altaffiltext{3}{MIT Kavli Institute for Astrophysics and Space Research, 70 Vassar Street, Cambridge, MA 02109, USA}
\altaffiltext{4}{MTA CSFK Lend\"ulet Near-Field Cosmology Research Group}
\altaffiltext{5}{ELTE E\"otv\"os Lor\'and University, Gothard Astrophysical Observatory, Szombathely, Hungary}
\altaffiltext{6}{MTA-ELTE Exoplanet Research Group, 9700 Szombathely, Szent Imre h. u. 112, Hungary}

\begin{abstract}
Compared with previous space-borne surveys, the Transiting Exoplanet Survey Satellite (TESS) provides a unique and new approach to observe Solar System objects. While its primary mission avoids the vicinity of the ecliptic plane by approximately six degrees, the scale height of the Solar System debris disk is large enough to place various small body populations in the field-of-view. In this paper we present the first data release of photometric analysis of TESS observations of small Solar System Bodies, focusing on the bright end of the observed main-belt asteroid and Jovian Trojan populations. This data release, named TSSYS-DR1, contains 9912 light curves obtained and extracted in a homogeneous manner, and triples the number of bodies with unambiguous fundamental rotation characteristics, namely where accurate periods and amplitudes are both reported. Our catalogue clearly shows that the number of bodies with long rotation periods are definitely underestimated by all previous ground-based surveys, by at least an order of magnitude. 
\end{abstract}


\section{Introduction}
\label{sec:introduction}

The Transiting Exoplanet Survey Satellite \citep[TESS]{ricker2015}
has successfully been launched on April 18, 2018 and after commissioning, started its routine operations on July 25, 2018. During the first two years of its primary mission, TESS observations are scheduled in terms of ``TESS sectors'' (or simply, sectors) where each sector corresponds to roughly 27 days of nearly continuous 
observations (in accordance with two orbits of TESS around Earth, with a spacecraft orbit in 1:2 mean-motion resonance with the Moon). 
The first year of observations ended on July 18, 2019, after completing the 13th sector (S13). Throughout these 13 sectors, TESS observed the primary fields on the Southern Ecliptic Hemisphere, covering the sky 
from the ecliptic latitude of $\beta$\,$\approx$\,-6$^\circ$, down to the southern ecliptic pole\footnote{\url{https://tess.mit.edu/observations/}}. 
This coverage is attained by four wide-field
cameras, each camera having a field-of-view (FoV) of 
$24^\circ\times24^\circ$ and the gross FoV is equivalent to a nearly contiguous rectangle in the sky, with a size of $96^\circ\times24^\circ$. 
The individual camera FoVs are also identified by the camera numbers and, according to the survey design, Camera \#4 continuously starred at the southern
ecliptic pole while Camera \#1 scanned the subsequent fields just south from the ecliptic plane. The cadence of TESS observations is $30$ minutes
in the so-called full-frame image (FFI) mode while pre-selected sources are observed with a cadence of $2$ minutes (hence, this mode is also called ``postage stamp'' mode). 
These two modes are also referred to as {\it long cadence} and {\it short cadence} observations: for TESS, long 
cadence also implies that the whole CCD frame is retrieved. 

This mission design allows us to observe Solar System objects during the primary mission, even considering the fact that the ecliptic plane is avoided by $\sim 6$ degrees. At first glance, objects with an
inclination higher than $\sim 6$ degrees are expected to be observed, but due to the $\sim$1\,AU distance of TESS to the Sun %finite ratio of TESS distance to the Sun 
and the semi-major 
axis range of $2.1-3.3$\,AU for the main-belt asteroids, also considering their non-zero eccentricities,
%as well as their eccentricity, 
thousands of objects with a few degrees of inclination are also possible to be observed with the aforementioned spacecraft attitude configuration. This limit of $6^\circ \lesssim i$ is more strict for distant objects, such as Centaurs or trans-Neptunian objects. 

According to earlier simulations \citep{pal2018}, one can expect good quality photometry of moving targets down to $V \lesssim 19\,{\rm mag}$ with a time resolution of $30$ minutes corresponding to the data acquisition cycle
of the TESS cameras in full-frame mode. Although the cadence for the postage stamp mode frames would allow a similar precision down to the
brighter objects (i.e. $V \lesssim 16\,{\rm mag}$), the corresponding pixel allocation would be too expensive. In this aspect, TESS short cadence observations are analogous the Kepler/K2 mission 
\citep{borucki2010,howell2014}
and similarly, only pre-selected objects could be observed in this mode \citep{szabo2015,pal2015}. Specifically, one should allocate roughly a thousand pixel-wise stamp if observations for a certain object are required. The rule-of-thumb for the apparent tracks of main-belt 
asteroids on long cadence TESS images is the movement 
of $\approx 1\,{\rm pixel}/{\rm cadence}$  \citep[see also Fig. 2 in][]{pal2018}. 
Of course, NEOs and trans-Neptunian objects could have apparent speeds which are larger and smaller, respectively. 

The yield of such a survey performed by TESS is a series of (nearly) 
uninterrupted, long-coverage light curves of Solar System objects -- 
like in the case of previous space-borne studies mentioned below.
% Ide jó lenne beidézni a korábbi K2-es cikkeket (SzR.)}). 
% ML: jönnek sorban, betettem rá utalást
From these light curves, one can obtain fundamental physical characteristics of the bodies
such as rotation periods, shape constraints and signs of rotating on a non-principal axis - with a much lesser ambiguity than in the case of 
ground-based surveys. This ambiguity is mainly due to the fact that ground-based photometric data acquisition is interrupted by diurnal variations -- 
which yield not just stronger frequency aliasing but higher fraction of long-term instrumental 
systematics. In addition, the knowledge of rotation period helps to resolve the ambiguity of rotation and thermal inertia \citep[see e.g.][]{Delbo2015} in thermal emission measurements of small bodies. 
%combining thermal measurements with rotational constraints can reveal many properties of the surface and hence the composition of the body. 
Further combination of spin information with thermal data  \citep[see e.g.][]{mueller2009,szakats2017,kiss2019}\footnote{\url{https://ird.konkoly.hu/data/SBNAF_IRDB_public_release_note_2019March29.pdf}} can therefore be an important initiative. 

This paper describes the first data release, TSSYS-DR1 of the TESS minor planet
observations, based on the publicly available TESS FFI data for the 
first full year of operations on the Southern Hemisphere. The structure
of this paper goes as follows. In the next section, 
Sec.~\ref{sec:objectselection} we describe how the objects were identified
and what kind of object selection principles are available for
a mission like TESS. In Sec.~\ref{sec:datareduction} we discuss the
main steps of the data reduction and photometry, emphasizing the importance
of differential image analysis. Sec.~\ref{sec:objects} summarizes the 
structure of the available data products while in Sec.~\ref{sec:comparison}
we make a series of comparisons with existing databases aiming to
collect photometric data series for small Solar System bodies. Our
findings are summarized in Sec.~\ref{sec:summary}.

\section{Object selection}
\label{sec:objectselection}

Regarding to the identification and querying Solar System objects on 
TESS FFIs, one can ask two types of questions: 
\begin{itemize}
    \item When and by which Camera/CCD was my target of interest observed?
    \item Which objects were observed by a certain Camera/CCD 
during a given sector?
\end{itemize}  
We can also connect these questions to the 
K2 Solar System observations. Namely, the first question is related to the 
computation of the pixel coverage of an asteroid track, as it 
was done in the case of K2 mission while observing pre-selected
objects \citep[see e.g.][]{pal2015,kiss2016,pal2016} and the second question
is related to the observations of serendipitous asteroids crossing
large, contiguous K2 superstamps \citep{szabo2016,molnar2018}.

In order to identify the objects which were observed by a certain Camera/CCD during a given sector, we followed a similar approach as it was done in our K2 asteroid studies \citep{molnar2018,szabo2016} and in the case of 
simulations of TESS observations \citep{pal2018}. Our solutions are based on an off-line tool called \texttt{EPHEMD}, providing a server-side
backend for massive queries optimized for defining longer time intervals 
and larger field-of-views within the same call 
\citep[see][for more details]{pal2018}. In fact, the catalogue presented
in this paper is retrieved by simply executing \texttt{EPHEMD} queries
on per-CCD basis for each sectors. Due to the dramatic decrease of the 
asteroid density at higher ecliptic latitudes, in this catalogue (DR1) 
we included only the observations from Camera \#1. 

\begin{table*}[!ht]
\caption{Quality flags and bits for the individual light curve data points.
The data point flags are interpreted in a bitwise logical-or combination of 
these individual flags. The bit positions between 0 and 11 
(values from 1 to 2048) are inherited from the FITS headers of the 
calibrated FFI data products, in accordance with the 
TESS Science Data Products Description Document\citep{tenenbaum2018}. 
The bit positions from 12 to 14 (mask values from 4096 to 16384) 
are specific for this particular data release and might be altered 
in the future. Note that bits at the position 1, 6, 8 and 9 
(having a description in parentheses) are not used in the TESS FFI 
data products.} 
\label{table:qualitybits}
\begin{center}\begin{tabular}{rrl}
Bit position & Value & Description \\
\hline
0 & 1 & Attitude Tweak. \\
1 & 2 & (Safe Mode.) \\
2 & 4 & Spacecraft is in Coarse Point. \\
3 & 8 & Spacecraft is in Earth Point. \\
4 & 16 & Argabrightening event. \\
5 & 32 & Reaction Wheel desaturation Event. \\
6 & 64 & (Cosmic Ray in Optimal Aperture pixel). \\
7 & 128 & Manual Exclude. The cadence was excluded because of an anomaly. \\
8 & 256 & (Discontinuity corrected between this cadence and the following one.) \\
9 & 512 & (Impulsive outlier removed before cotrending.) \\
10 & 1024 & Cosmic ray detected on collateral pixel row or column. \\
11 & 2048 & Stray light from Earth or Moon in camera FOV. \\
\hline
12 & 4096 & Formal photometric noise exceeds the threshold of $0.5$ magnitude. \\
13 & 8192 & Point rejected due to the presence of unexpected histogram region. \\
14 & 16384 & Manual removal of an outlier point. 
\end{tabular}\end{center}
\end{table*}

\newlength{\cheight}
\setlength{\cheight}{27mm}
\newcommand{\rcentered}[1]{\rotatebox{90}{\begin{minipage}{\cheight}\centering#1\end{minipage}}}

\begin{figure*}[!ht]
\begin{center}
\begin{tabular}{cccc}
\rcentered{Original} 				& \resizebox{!}{\cheight}{\includegraphics{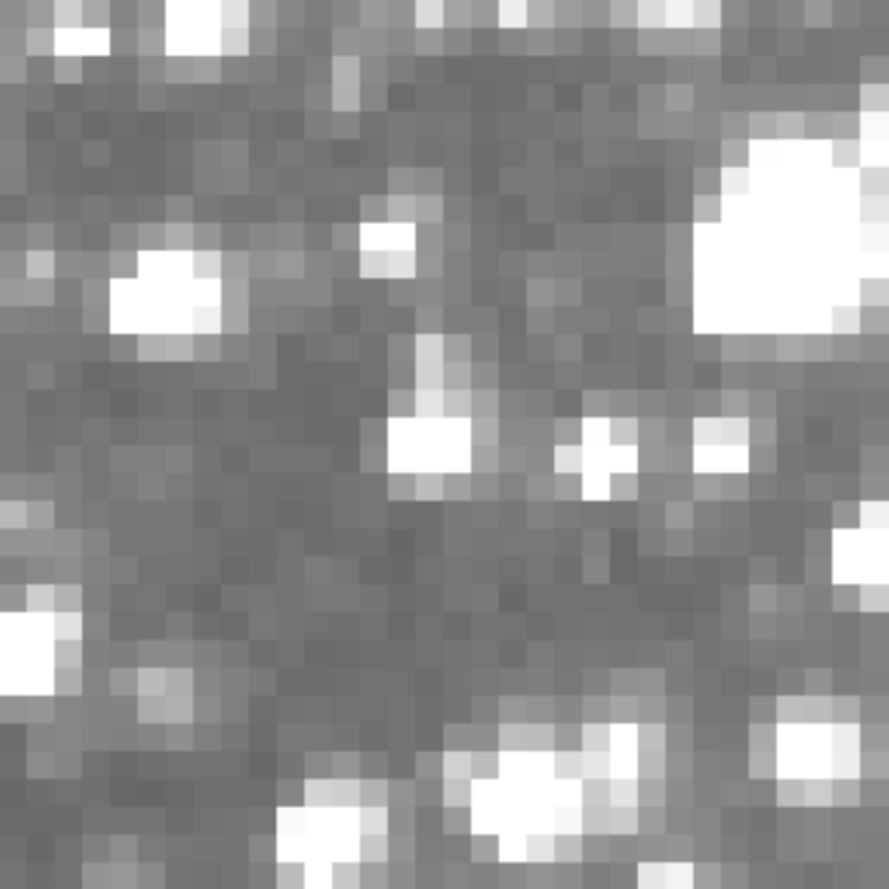}}&\resizebox{!}{\cheight}{\includegraphics{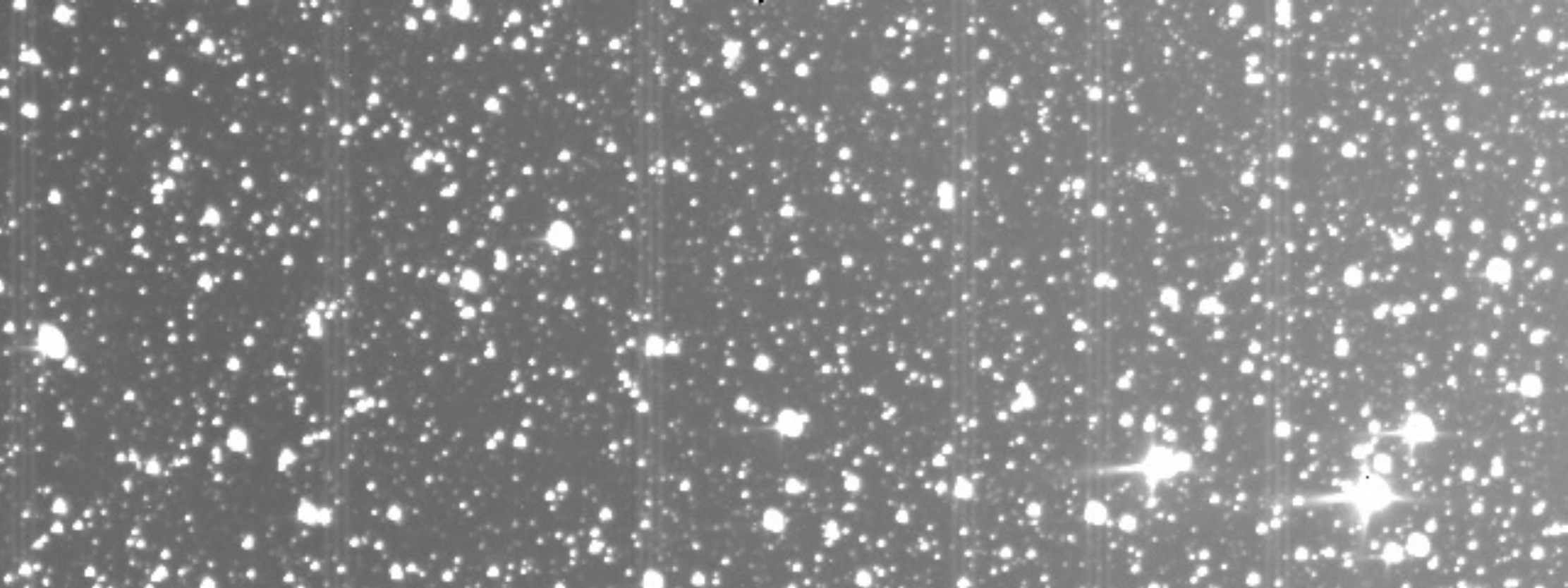}}&\resizebox{!}{\cheight}{\includegraphics{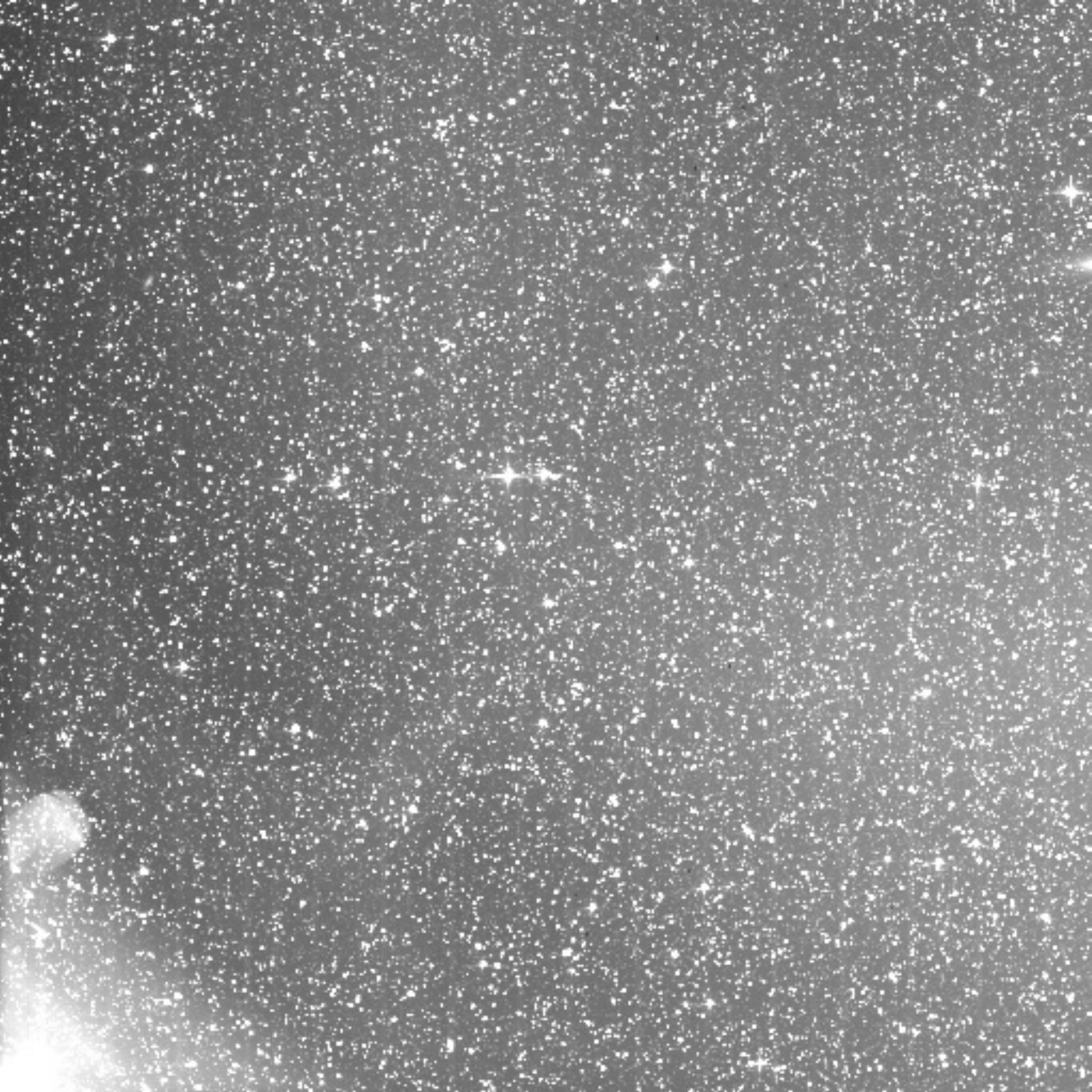}} \\
\rcentered{Background difference} 		& \resizebox{!}{\cheight}{\includegraphics{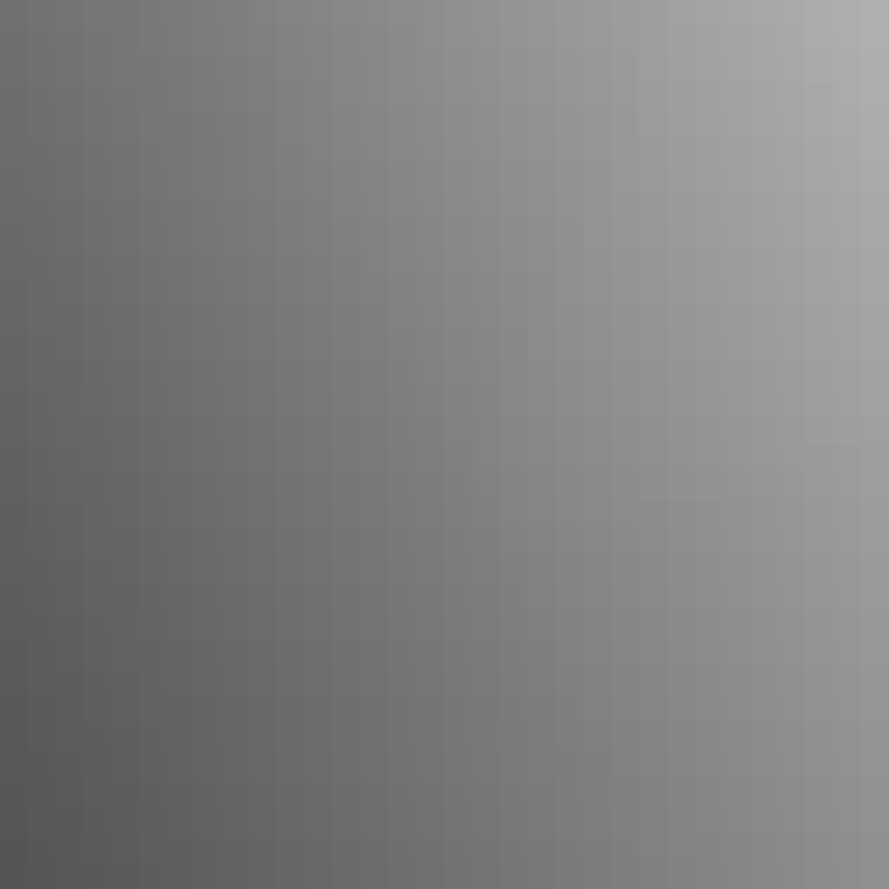}}&\resizebox{!}{\cheight}{\includegraphics{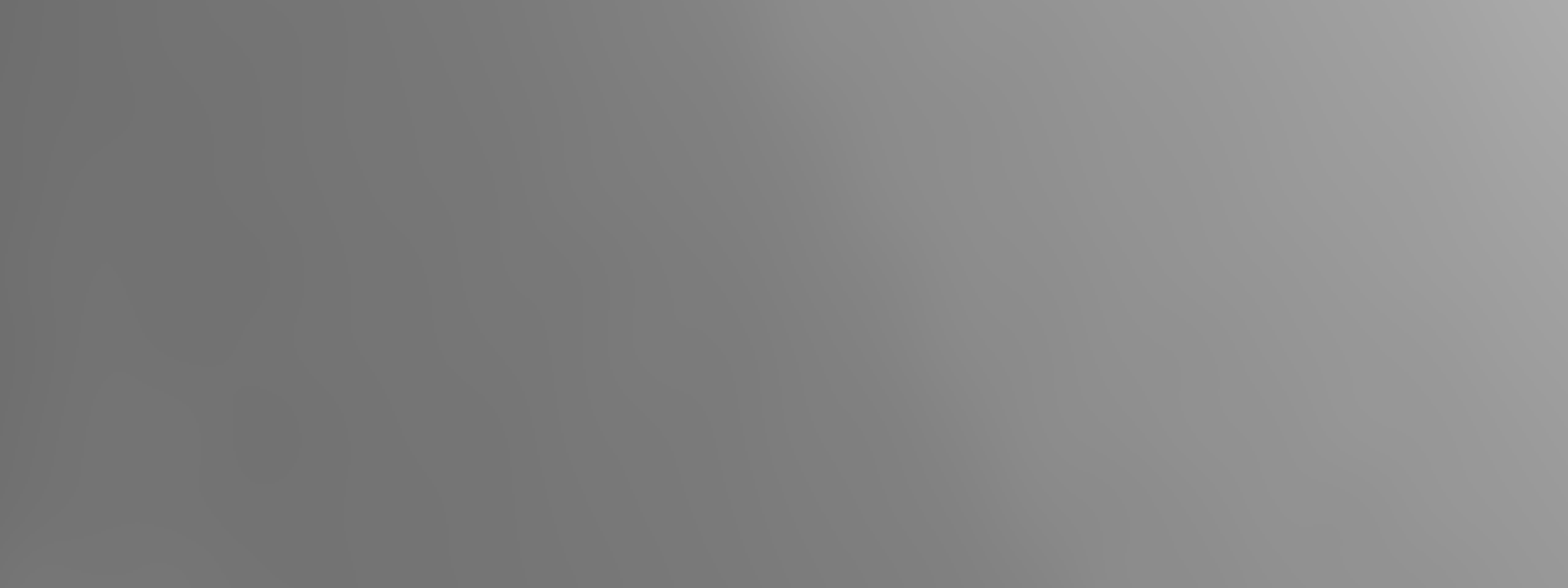}}&\resizebox{!}{\cheight}{\includegraphics{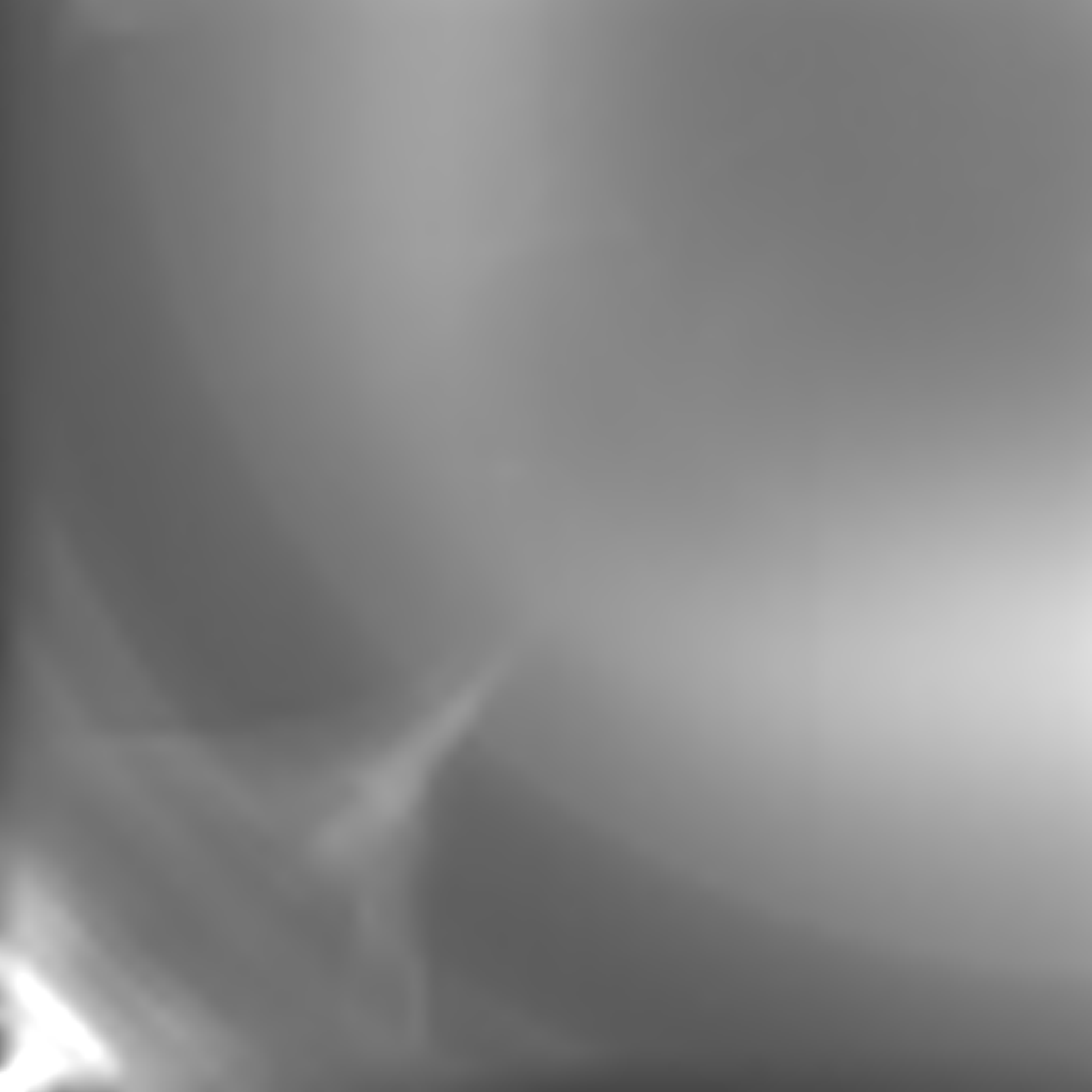}} \\
\rcentered{Background variations removed}	& \resizebox{!}{\cheight}{\includegraphics{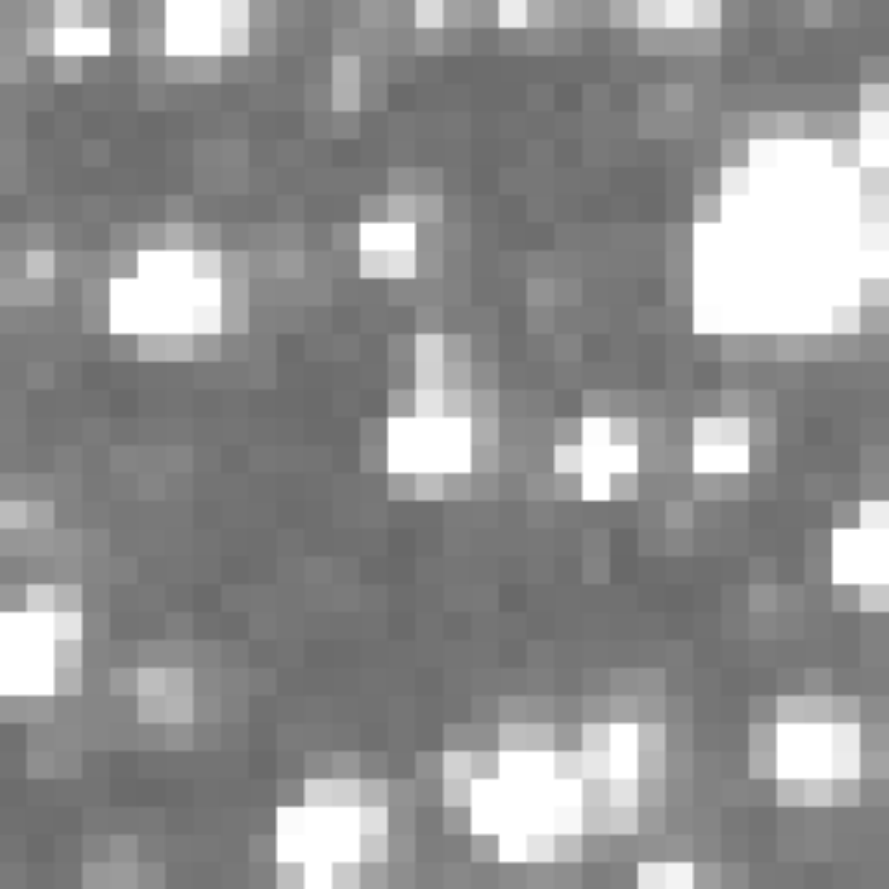}}&\resizebox{!}{\cheight}{\includegraphics{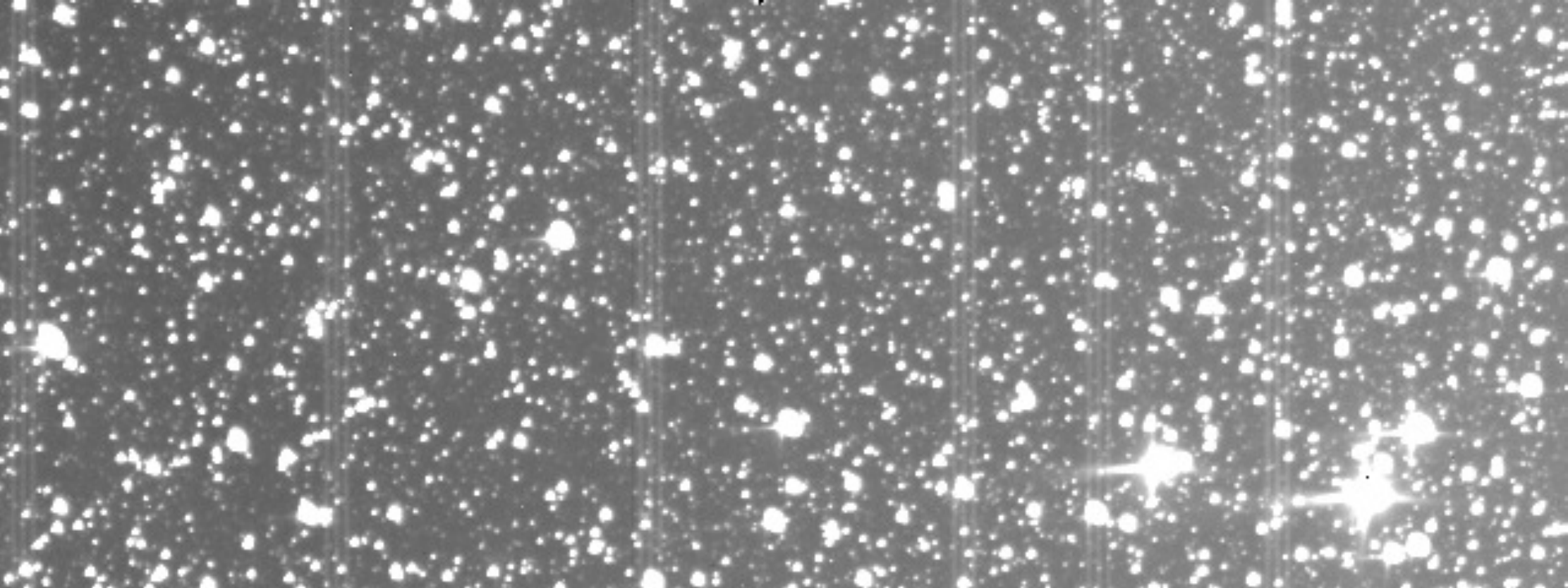}}&\resizebox{!}{\cheight}{\includegraphics{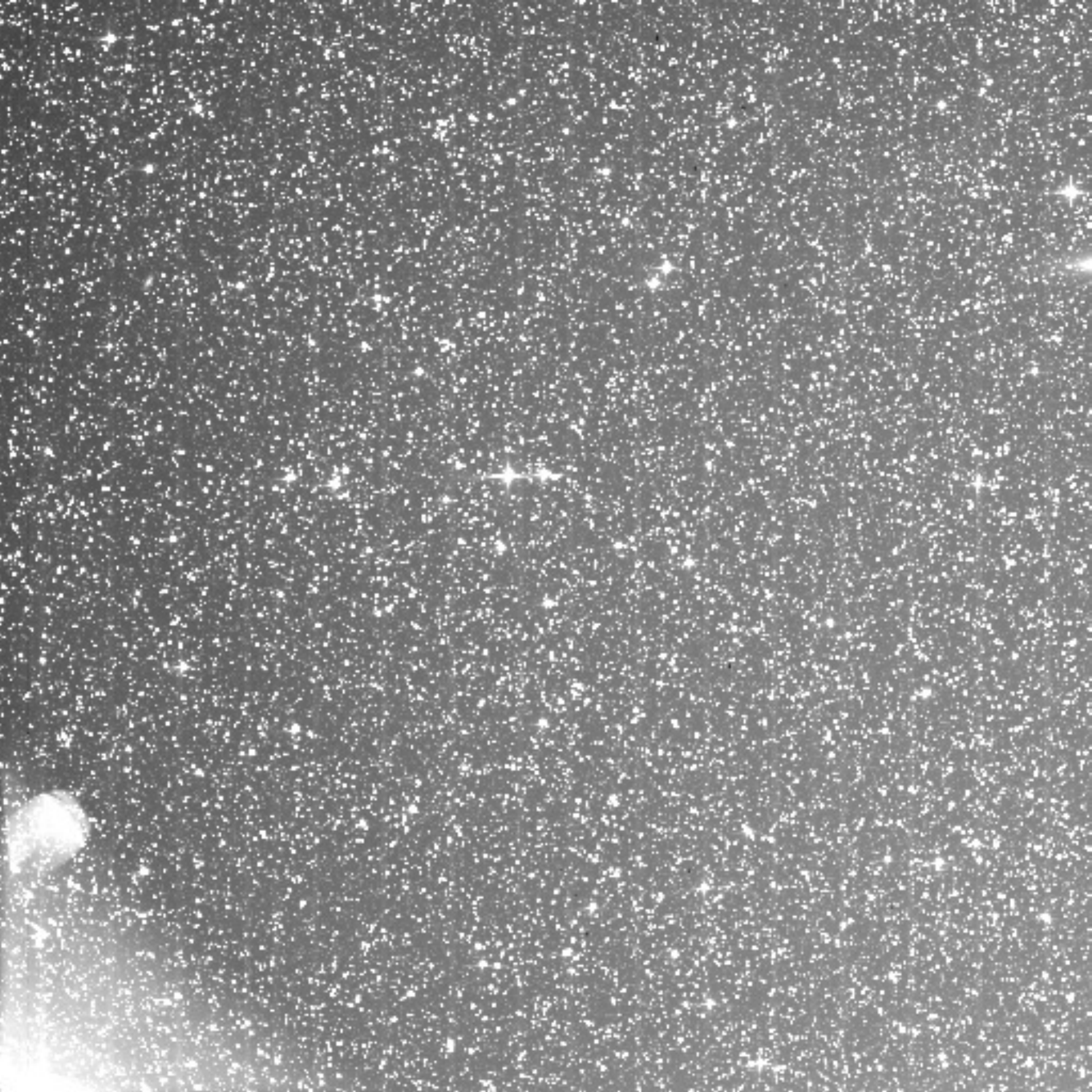}} \\
\rcentered{Simple difference} 			& \resizebox{!}{\cheight}{\includegraphics{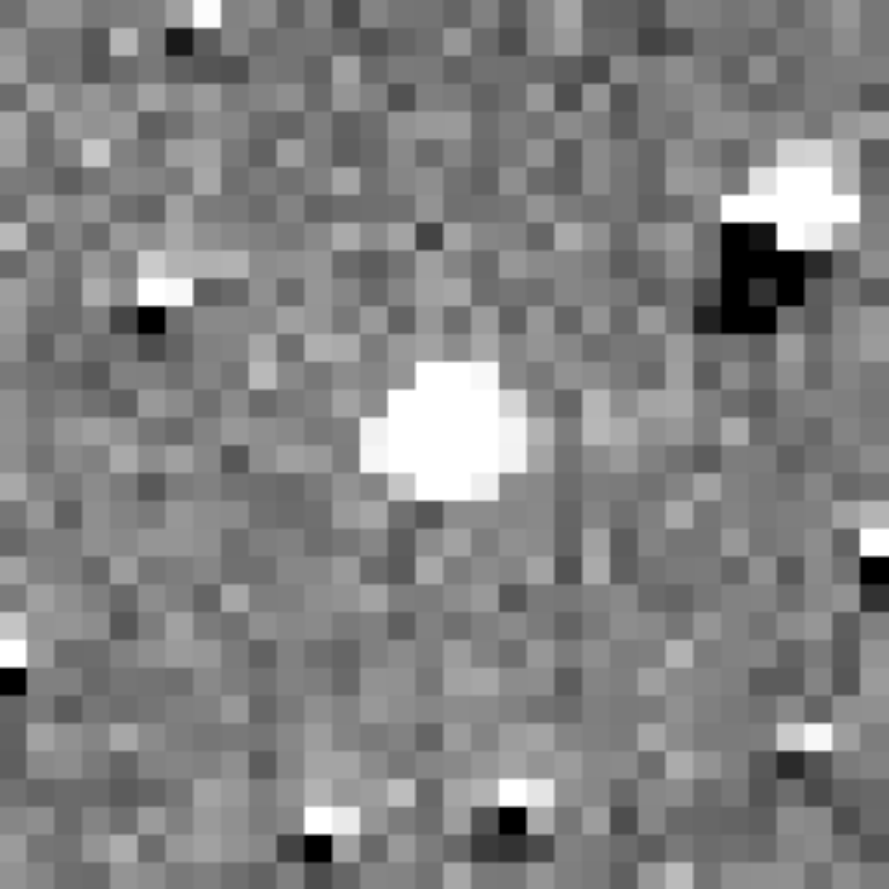}}&\resizebox{!}{\cheight}{\includegraphics{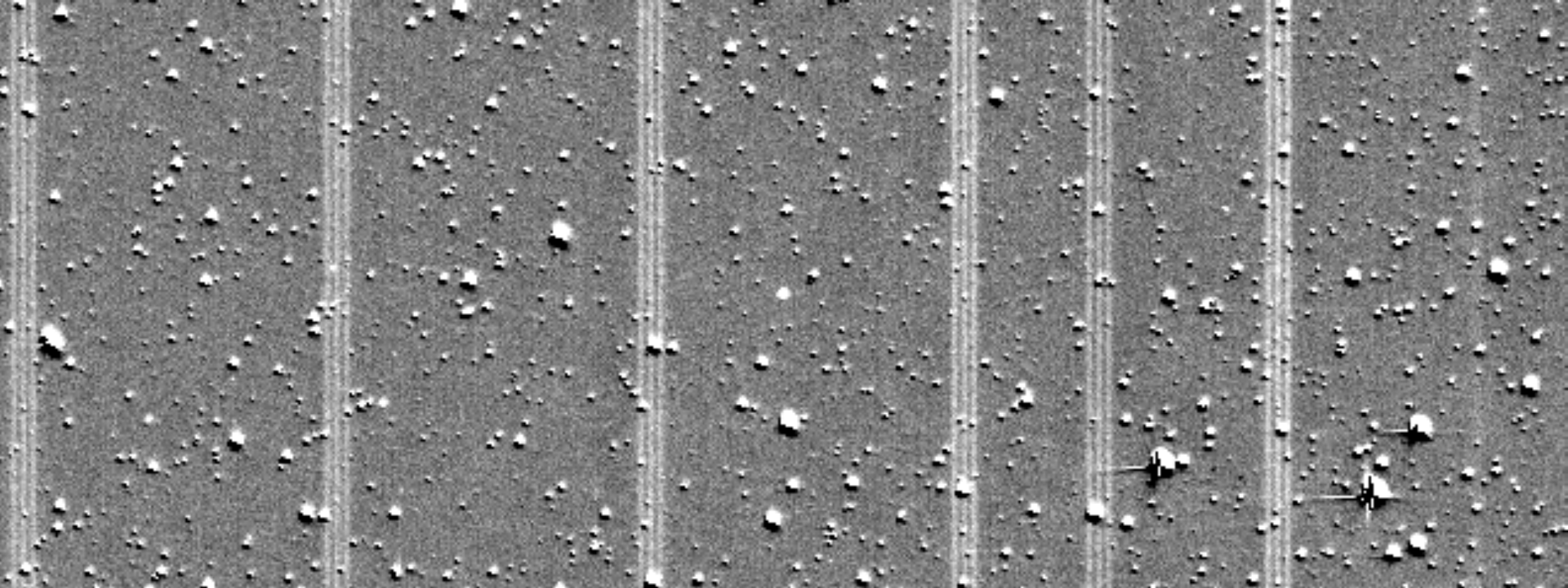}}&\resizebox{!}{\cheight}{\includegraphics{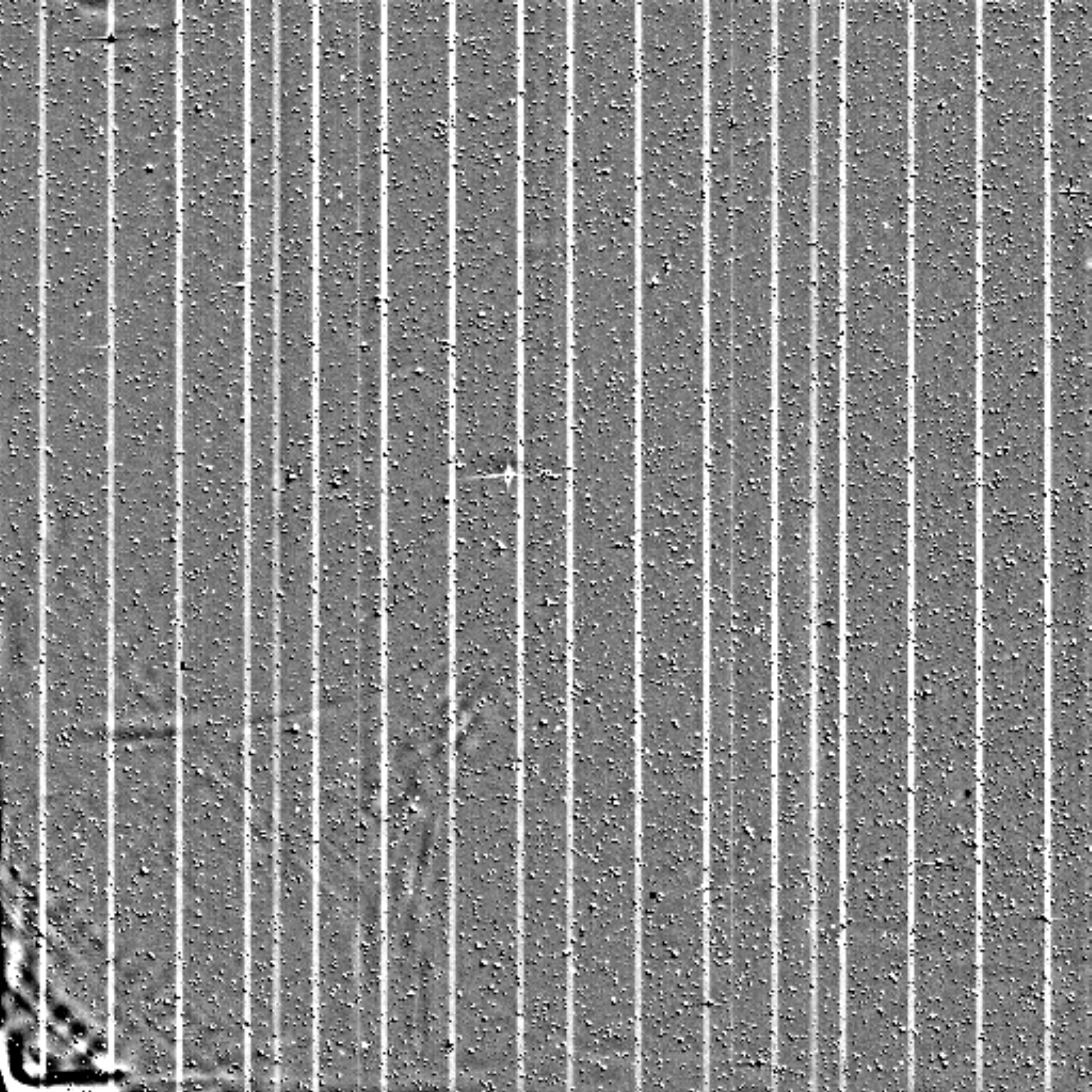}} \\
\rcentered{Convolution \& difference} 		& \resizebox{!}{\cheight}{\includegraphics{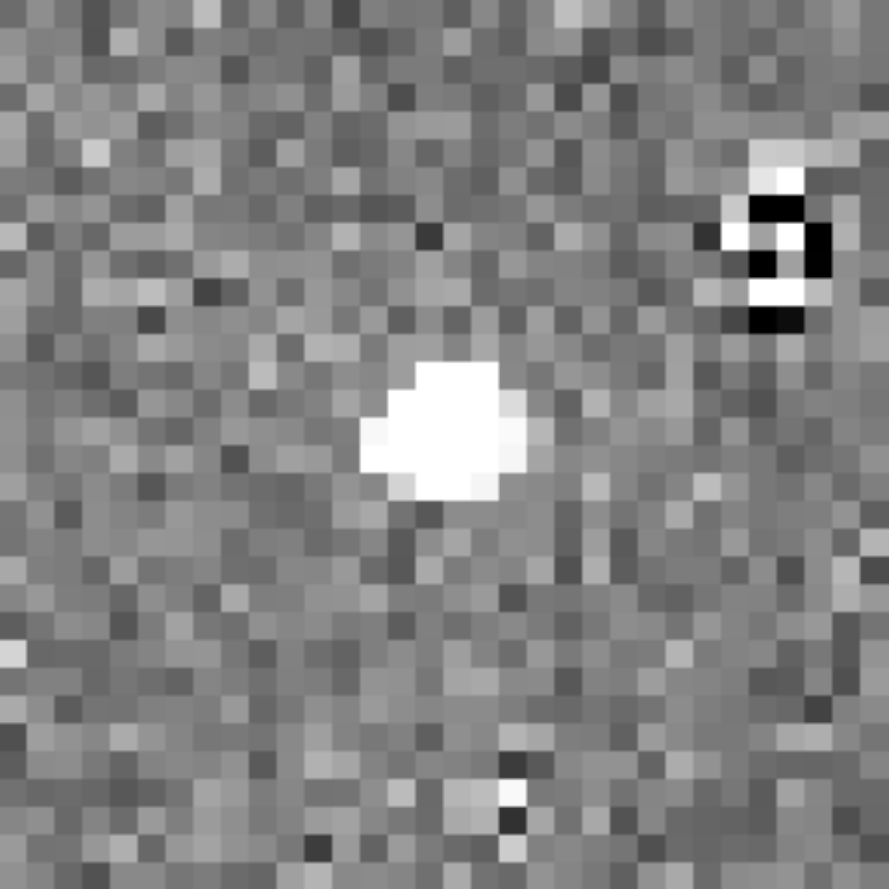}}&\resizebox{!}{\cheight}{\includegraphics{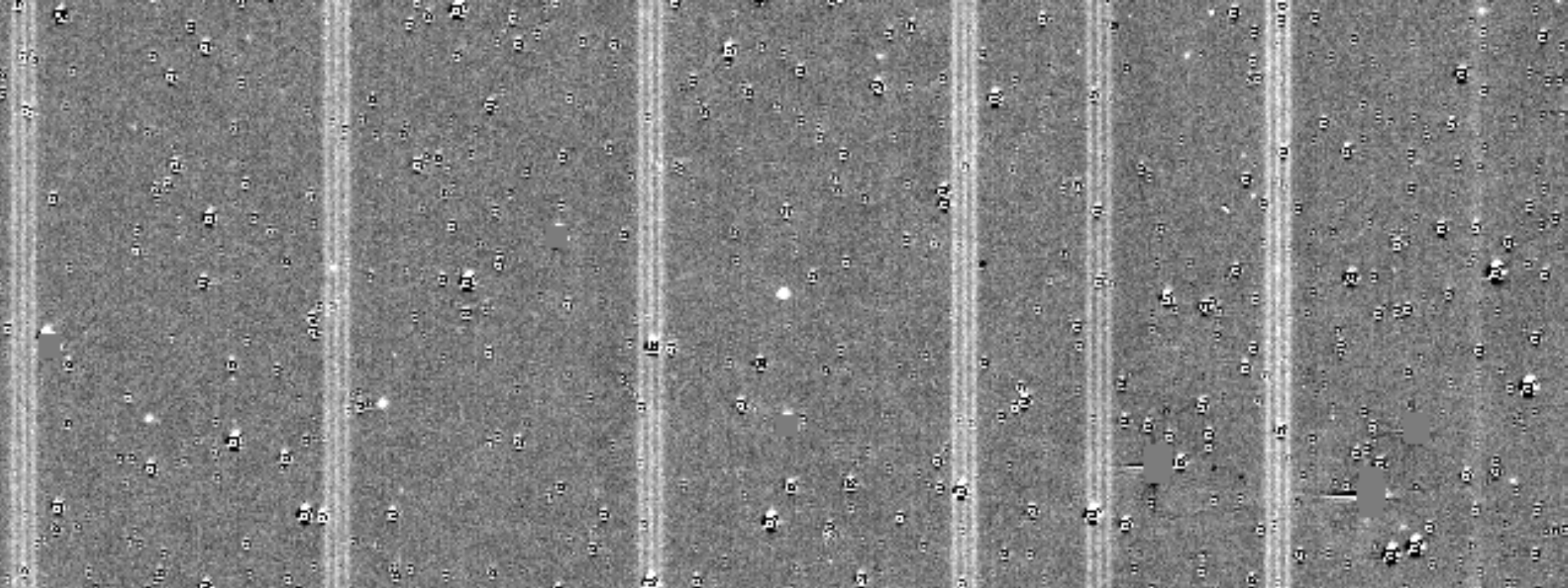}}&\resizebox{!}{\cheight}{\includegraphics{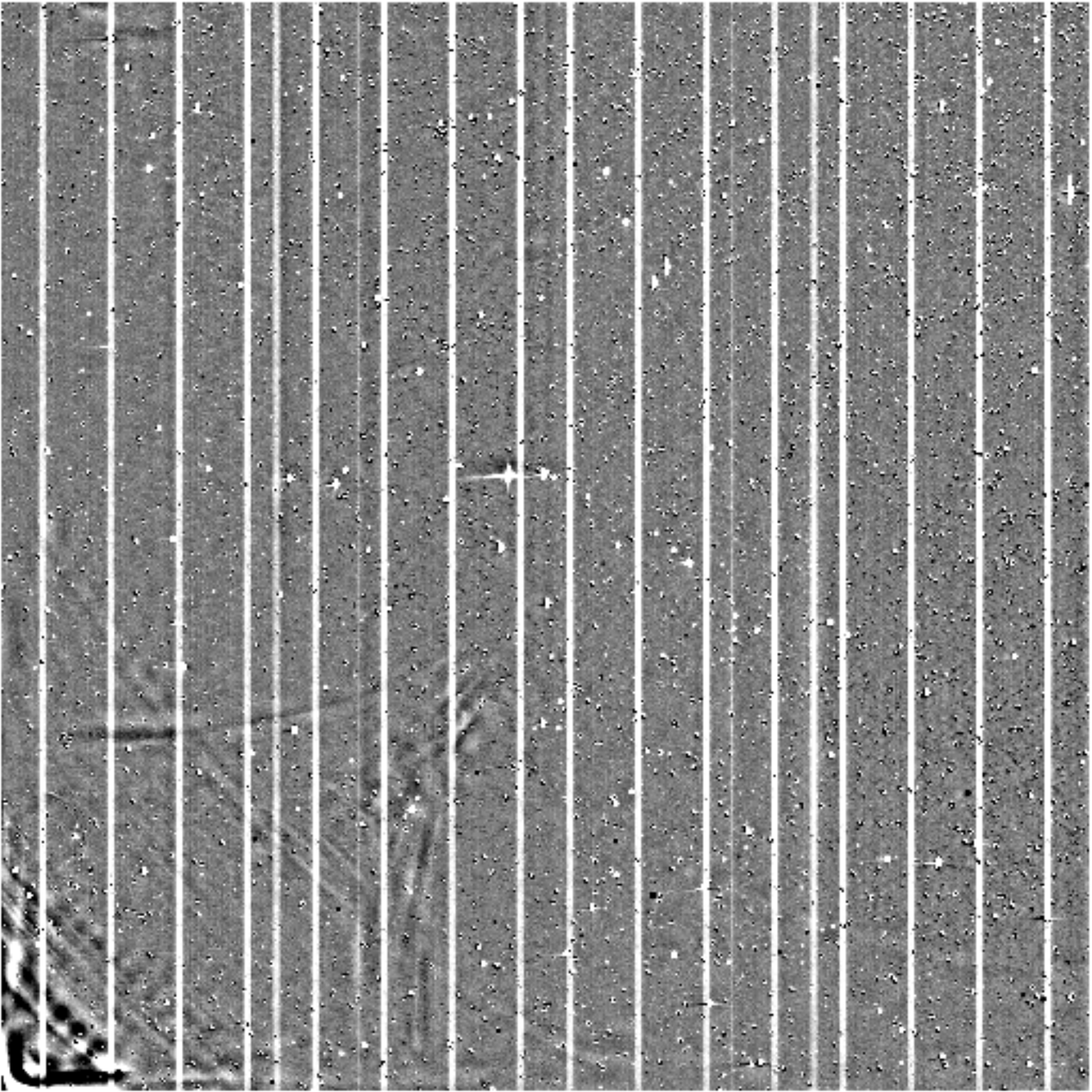}} \\
\rcentered{Stripes removed} 			& \resizebox{!}{\cheight}{\includegraphics{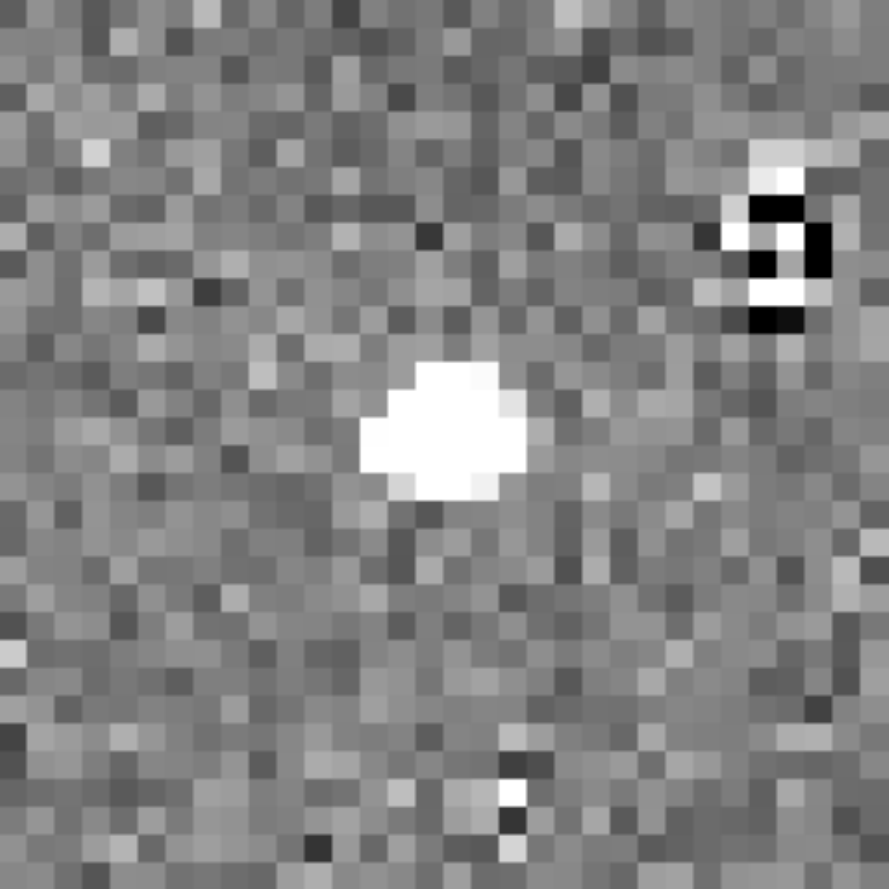}}&\resizebox{!}{\cheight}{\includegraphics{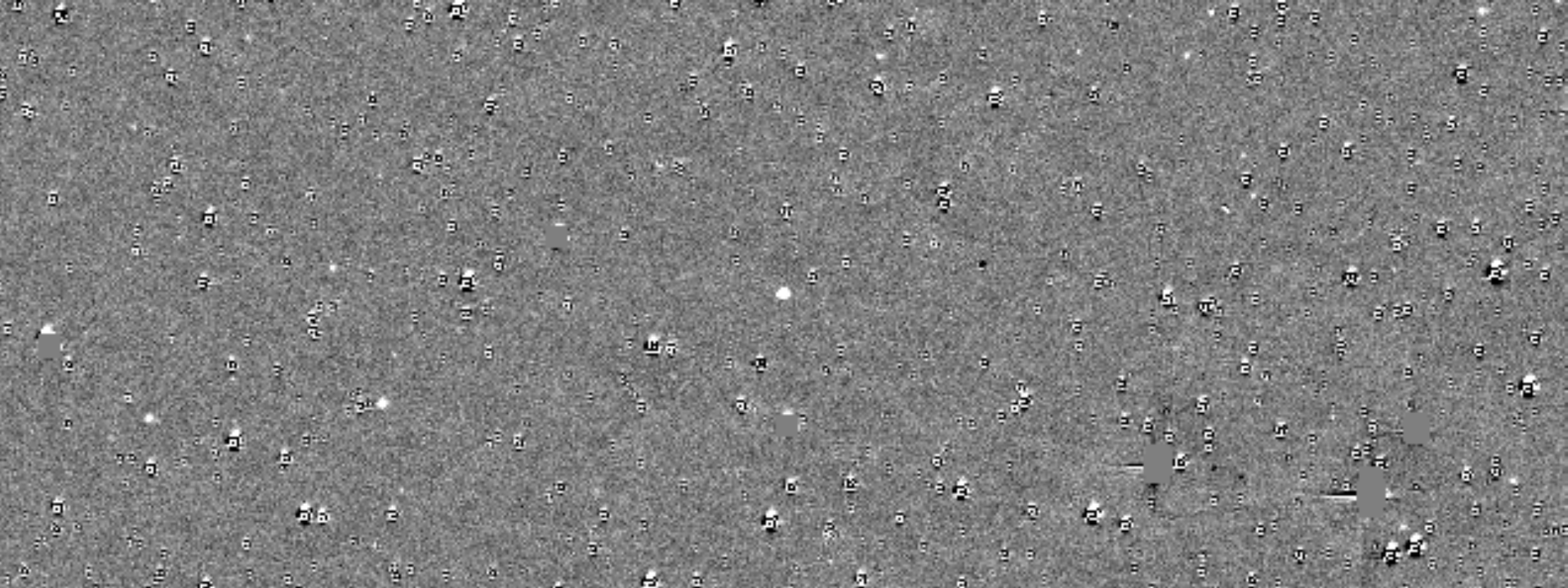}}&\resizebox{!}{\cheight}{\includegraphics{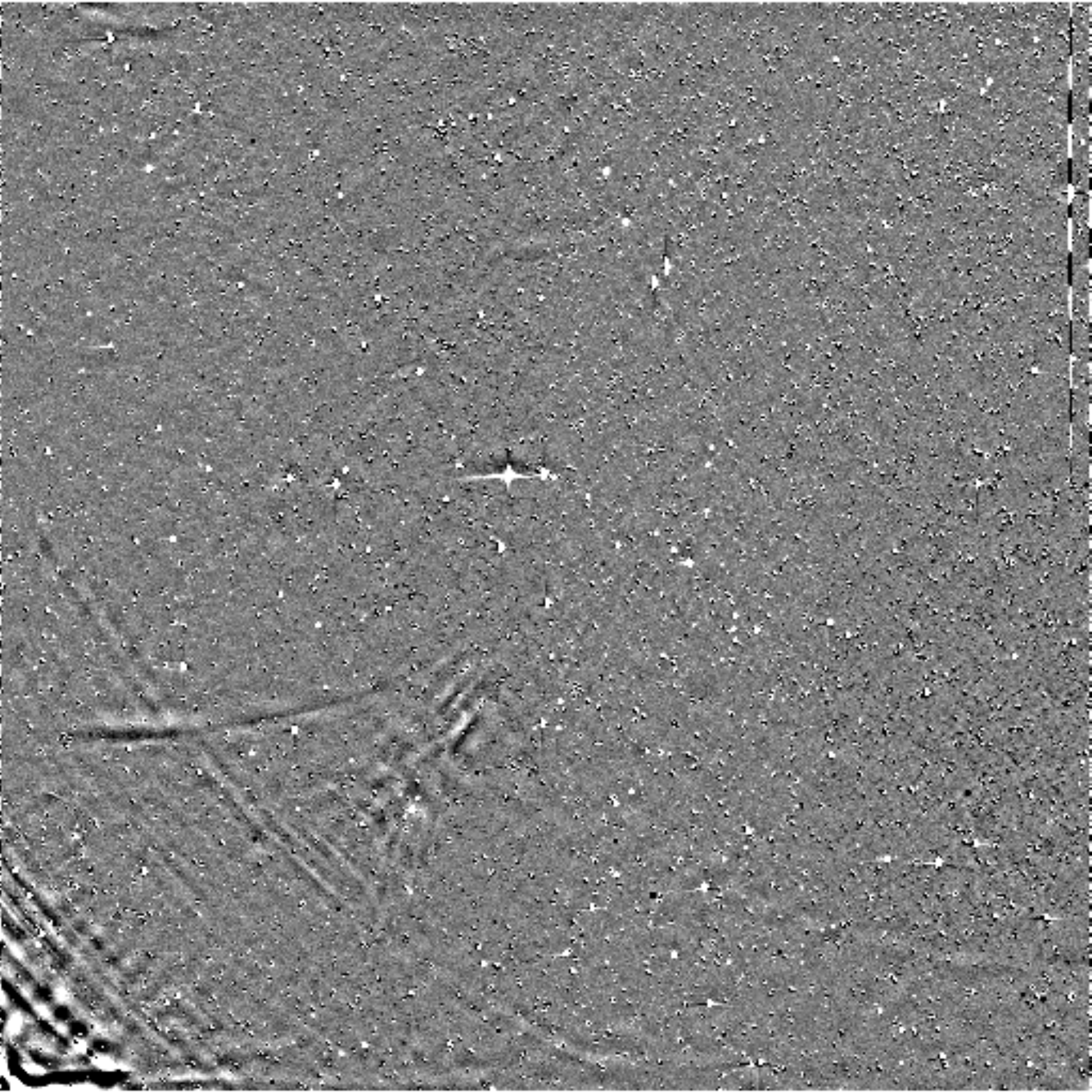}}
\end{tabular}
\end{center}
\caption{Panels showing the various stages of the image-level data 
processing using asteroid (2429) Schurer as an example, observed 
during Sector 2, by Camera \#1, CCD \#3. The {\it left} column shows the
$10^\prime\times10^\prime$ vicinity of the target, the {\it middle}
column shows the neighbourhood ($3.7^\circ\times1.4^\circ$) area
while the {\it right} column is the full CCD frame,
all at the TESS FFI cadence 2018247095941 (JD 2458365.92767).
Images in the {\it first row} show the original unprocessed data.
The {\it second row} is the difference in the background structure
with respect to a frame where the Earth and Moon were below the
sun-shade of TESS. The large-scale variations due to the stray light
are clearly visible. 
The {\it third row} shows the difference between
the first two rows. 
The {\it fourth row} shows the naive difference between the target
image and the median differential-background reference image. 
The residuals due to the uncorrected differential
velocity aberration are clearly visible.
The {\it fifth row} shows the results of the
image convolution followed by subtraction. This step also makes the 
TESS-specific, but otherwise comparatively faint vertical CCD stripes visible. 
In addition, the left stamp in this row shows that the sources, even ones
brighter than the target objects are completely removed, with some residual
structure only visible at a much brighter star at the upper-right corner
of this stamp.
Images in the {\it sixth row} show the results of the stripe removal
process. The target at the center is clearly visible.}
\label{fig:datared}
\end{figure*}

\section{Data reduction and photometry}
\label{sec:datareduction}

As it was mentioned above, the whole data processing of this catalogue
was based on the observations performed by Camera \#1 while surveying 
TESS sectors ranging from 1 up to 13. The processing has been carried out
on a per-CCD basis, executing the same set of routines on the 
$13\times 4=52$ blocks of images corresponding to a single-sector-single-CCD
acquisition run.  The pipeline providing the light curves is exclusively based on the FITSH package \citep{pal2012}. In this section we summarize the main steps of the photometric processing.

\subsection{CCD-level steps}
\label{sec:ccdlevelprocessing}

Each of the CCD image series is processed as follows. Based on the 
available orbital and pointing data, we selected nearly a dozen of frames
called {\it individual median reference frames} (IMRFs)
spanning a $\sim 2$-day period long interval close to the center of the
observations evenly. These frames coincide for all of the four CCDs for a given
sector, i.e., these correspond to the same cadence and usually have a time
step of 4 hours between each frame. Another set of criteria was based on the
constraint that both the Sun and the Moon should have been below the 
sun-shade of the spacecraft, meaning that both the Sun-TESS-boresight and the 
Moon-TESS-boresight angle should have been larger than $90^\circ$. This
combined selection criteria ensured the lack of stray light in {\it all}
of the cameras at the same time while the duration ensured an expected
coverage of several tens of pixels of a main-belt asteroid while still
keeping the differential velocity aberration at a considerably low level. 
In addition to the aforementioned selection criteria, if 
a prospective frame was flagged with a ``reaction wheel desaturation event''
\citep[see][]{tenenbaum2018}, the next or previous frame was selected instead. 

In the next step, IMRFs were used to create a median image, employed
as a {\it median differential background reference image} (MDBRI). 
This MDBRI was then subtracted from all of the images acquired by the same
CCD in the same sector and the resulting differences were smoothed using
a median window filtering combined with spline interpolation with a grid
size of $64\times 64$ pixels. This step allowed the derivation of 
large-scale background variations and nicely helped to minimize and model
the variations inducted by scattered light and zodiacal light. 
The derived background variations were then subtracted from all of the 
images and image convolution were applied between the MDBRI and these 
background-subtracted images. Note that this step does not subtract 
the {\it intrinsic} background since such a background practically does not
exist for TESS images due to the very strong confusion and large pixel size. 
The image convolution steps correct not only for the PSF variations but for the
offsets inducted by the differential velocity aberration as well. The latter
one can be as large as one tenth of a pixel throughout a sector and it is
the most prominent further away from the spacecraft boresight (which includes
Camera \#1 CCDs \#3 and \#4, which are the closest to the ecliptic plane).
Once the convolved MDBRIs are derived, the resulting residual image
was processed by a spline-smoothed median window filtering with a 
block size of $1\times 64$ pixels. This filtering removed the vertical
stripes exposed in the TESS CCDs in parallel with the increased stray light.
The steps of the aforementioned processing are displayed in Fig.~\ref{fig:datared} via the example of (2429) Schurer. 

\begin{figure}[!ht]
\plotone{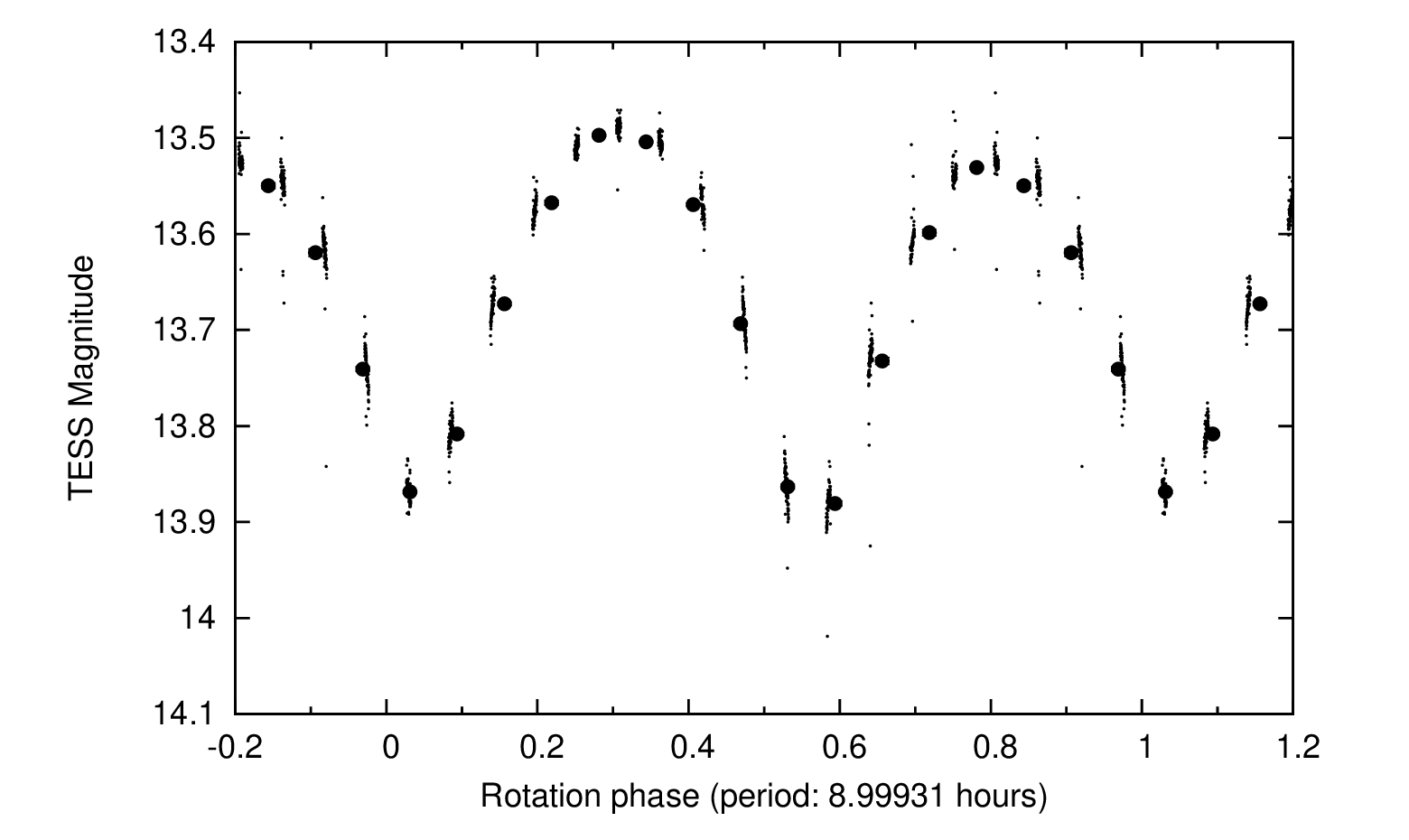}
\caption{Folded light curve of (692) Hippodamia, having a rotation period 
$P=8.9993\,{\rm h}$. While this rotation period satisfies the Nyquist criterion, the phase coverage is not uniform due to the $P/C$ ratio of $\sim 18$.}
\label{fig:hippodamia}
\end{figure}

%\begin{figure*}[!ht]
%\hrule
%\plotone{img/pv_354}
%\hrule
%\plotone{img/pv_220281}
%\hrule
%\caption{Validation plots for (354) Eleonora (top four graphs on a 
%single sheet) and (220281) 2003\,BA$_{47}$ (bottom four graphs on 
%a single sheet). These plots are available for all of the 9912 objects 
%presented in this study.}
%\label{fig:validationplots}
%\end{figure*}

\begin{figure*}[!ht]
\plottwo{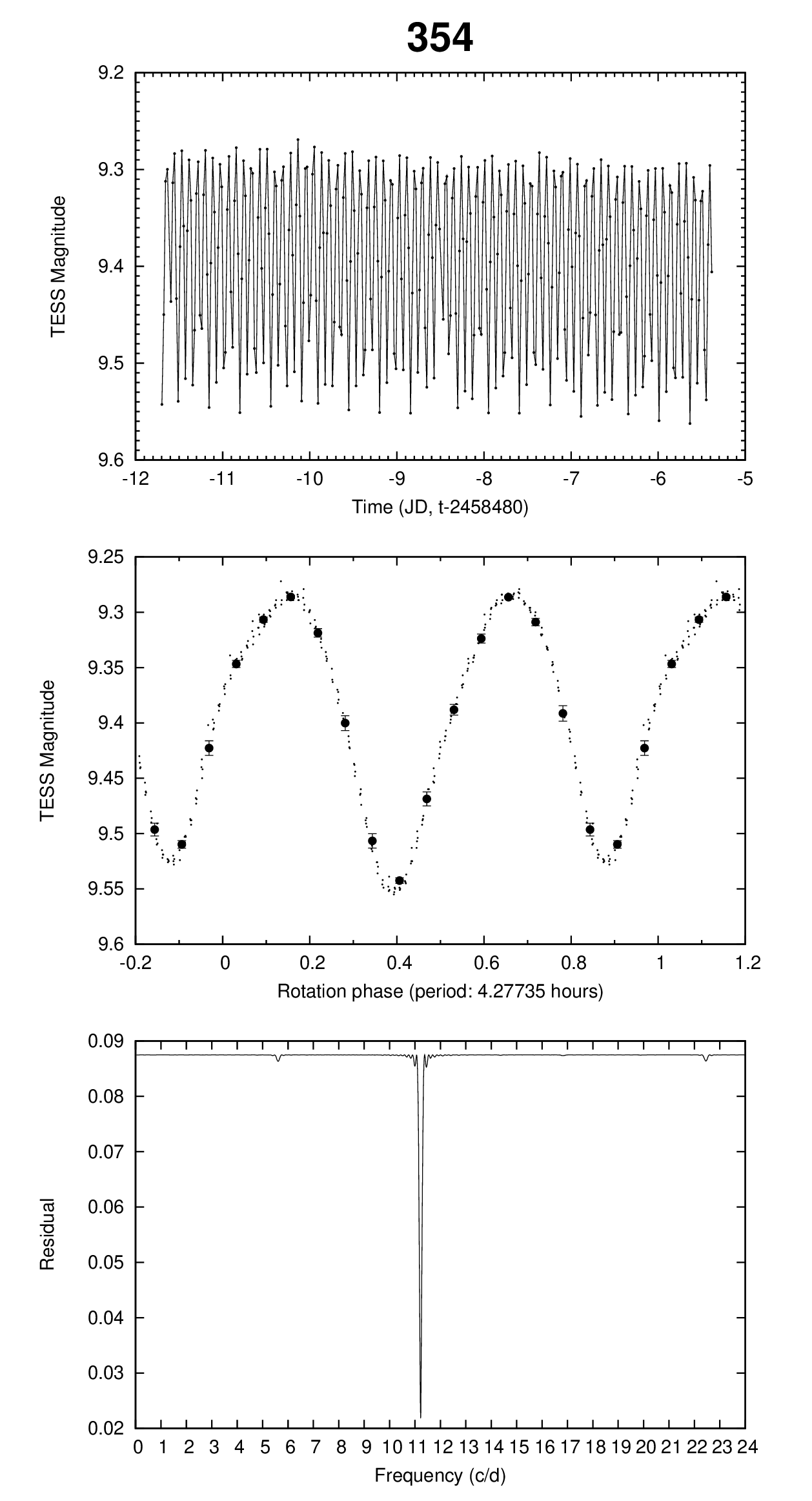}{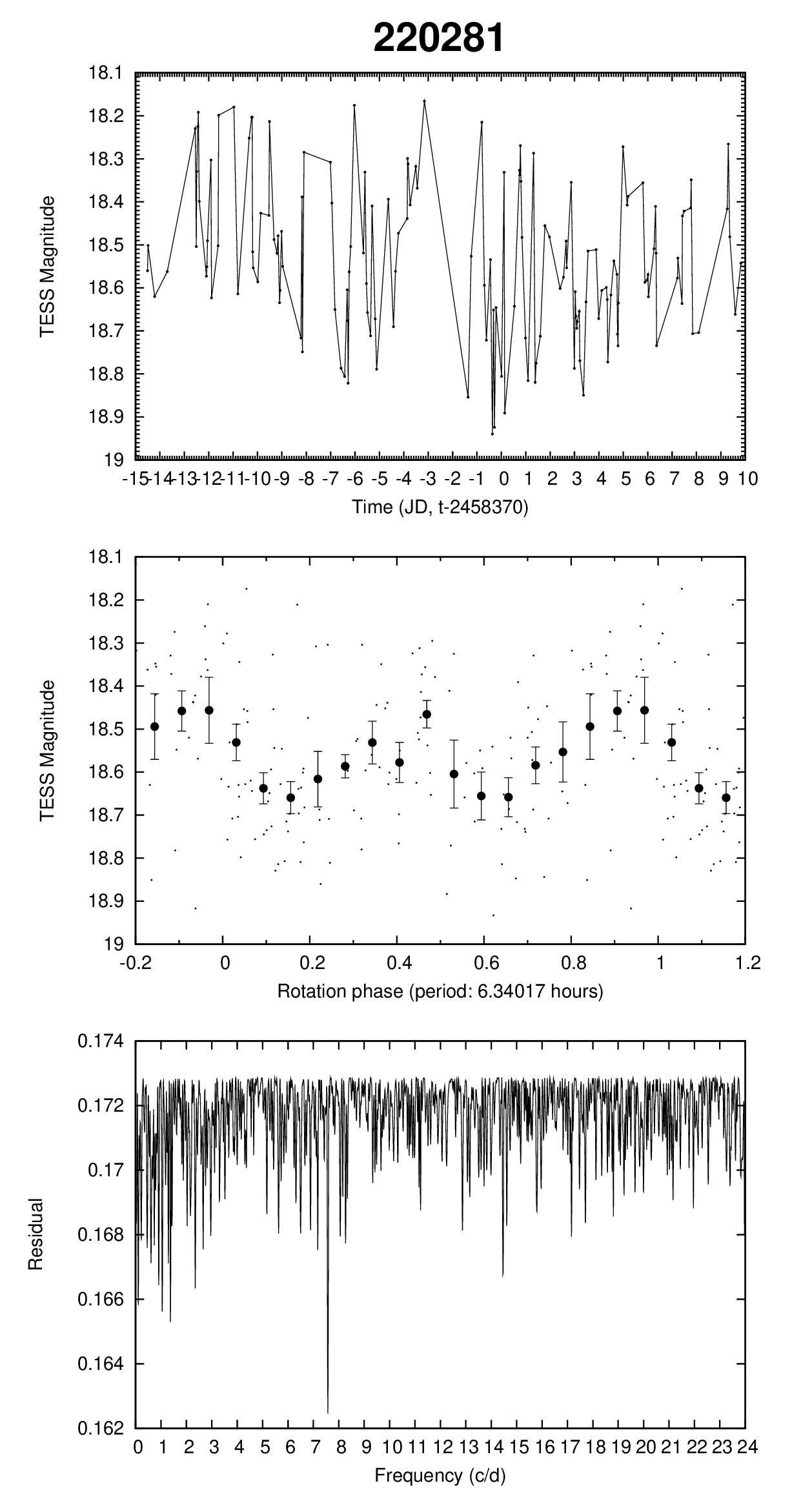}
\caption{Object light curve plots for (354) Eleonora (left column of 
3 individualplots) and (220281) 2003\,BA$_{47}$ (right column of 3 
individual plots). These plots are available for all of the 9912 objects 
presented in this study.}
\label{fig:objectplots}
\end{figure*}

\subsection{Target astrometry and photometry}
\label{sec:targetphotometry}

These cleared images were then used as the input of the aperture photometry where the centroids are computed by the \texttt{EPHEMD} tool with TESS set as the observer's location. Absolute astrometric plate solutions have been derived using the \textit{Gaia} DR2 catalogue \citep{gaia2016,gaia2018} while the projection function was obtained by a third-order Brown-Conrady model on the top of tangential projection with additional refinements using a third-order polynomial expansion. The fluxes are extracted using the proper way needed to interpret convolved differential images \citep[see Eq. 83 in][]{pal2009}. 
{The zero-point of the light curves were obtained using a global fit against the GAIA DR2 RP magnitudes. 
Due to the almost perfect overlap of the TESS and GAIA RP passbands -- see also Fig.~1 in 
\cite{ricker2015} and Fig.~3 in \cite{jordi2010} -- 
this yields a good and accurate match of the zero point. 
However, offsets can be presented due to the PSF variations 
across the field-of-view of the fast TESS optics. We note here 
that the formal uncertainties does not include the respective uncertainty of this offset.
Individual light curve files were then
generated by transposing the photometric results and flagged afterwards
according to the quality flags presented in the TESS FFI headers
\citep{tenenbaum2018}. Light curves with insufficient number of data points
were removed from the database and the post-filtering of these remaining
light curves also added additional types of quality flags 
(see Table~\ref{table:qualitybits}). This post-filtering process includes
exclusion of the points with high formal photometric uncertainty,
outlier detection based on histogram clipping and manual removal of points
in the most prominent cases. 

The filtered light curves were then analyzed by performing a period search.
This period search was based on fitting a sinusoidal variation in 
parallel with the decorrelation of the phase angle variations up to 
the second order (see also Sec.~\ref{sec:spectrum} later on).
The most dominant frequency was computed by interpolating
in the vicinity of the frequency spectrum were the root mean square of the aforementioned fit residual was found to be the smallest (see Section \ref{sec:spectrum}). The light curves were then folded and binned after phase angle correction. Folding was performed with two periods, one corresponding to the dominant frequency while the other period we used was twice the dominant period, assuming a double-peaked light-curve generated by the rotation of an elongated body. 

In total, 9912 objects are included in the present data release, for which
accurate light curve information were derived with a reasonable significance.
Out of these 9912 objects, 125 have only provisional designations and 
therefore are not numbered minor planets.

\subsection{Sampling characteristics}
\label{sec:sampling}

The observing strategy of TESS is highly deterministic 
compared to many of the surveys and ground-based observations. 
Namely, the cadence is 
strictly $C=0.5\,{\rm h}$ for a nearly uninterrupted 
observing period of $L\lesssim25-28\,{\rm d}$. This property implies the Nyquist criterion 
which does not allow the unambiguous rotation characterization for objects having 
a period of $P\leq 2C=1\,{\rm h}$. This is interesting for small objects, having 
a size of approximately or smaller than the spin barrier limit of $\sim 100\,{\rm m}$: such objects 
can rotate faster than $\sim 2.2\,{\rm h}$ \citep{pravec2000}.

The strict cadence also yields sampling artefacts of objects having a 
rotation period which is close to the integer multiple of the cadence $C$. 
For instance, (692) Hippodamia has a rotation period of $P=8.9993\,{\rm hours}$, 
which is almost exactly $18$ times longer than the TESS FFI 
cadence (see Fig.~\ref{fig:hippodamia}). In order to 
characterize the strength of this sampling effect, 
let us assume that the period of the object is $P=nC+\varepsilon$ where $\varepsilon$ 
represent a short time difference and $n$ is an integer 
number (e.g. $n=18$ and $\varepsilon=-0.0007\,{\rm h}$ for (692) Hippodamia). 
In order to fully sample the rotational phase domain, one should expect that the second 
instance ($t=C$) has the same phase as the last phase after at or around the $k$th 
rotation where for the total observation timespan is $L\approx kP$. Here $k$ is also 
an integer, the total number of rotations covered during the observations. The phases 
are equal if $(knC/P)-(C/P)=k$, from which we can compute that $CP/L$ should be 
smaller than $|\varepsilon|$. This limit for (692) 
Hippodamia is $|\varepsilon|_{692}=(1/2\,{\rm h})\cdot(8.9993\,{\rm h})/(25\,{\rm d})\approx0.0075\,{\rm h}$, 
definitely larger than $|\varepsilon|=0.0007\,{\rm h}$, we obtained above for this object, 
resulting in a stroboscopic effect. This stroboscopic effect is also present in K2 
observations, see e.g. the case of (14791) Atreus in \cite{szabo2017}.

\section{Database products and structures}
\label{sec:objects}

Per-object data products were saved and stored in accordance with the
aforementioned steps. The primary data products include four files 
per object, namely:
\begin{itemize}
\item the light curve file, containing the time series of the brightness
measurements for a particular object;
\item the residual r.m.s.\ frequency spectrum;
\item a metadata file (best-fit rotation frequency, peak-to-peak amplitude, light curve type); and
\item validation sheets, including the plots of the aforementioned data products,
\item and per-object summary plots and slides, including the folded light curve with the most likely rotation period.
\end{itemize}
In the following, we describe these data products in more detail. The full data release is going to be available from the web address of \url{http://archive.konkoly.hu/pub/tssys/dr1/}.

\subsection{Light curve files}
\label{sec:lcfiles}

The light curve files basically represent the post-transposition stage of the 
photometric output. Since photometry is performed on a per-frame basis
and a single call to the photometric task (FITSH/\texttt{fiphot}) performs
the flux extraction for all of the minor planets associated with that
particular frame, light curve files also include the target name, the 
timestamp, the $(x,y)$ pixel coordinates and estimations for the background 
structure. Although differential
imaging analysis and the subsequent photometry yields zero local background 
on subtracted images in theory, some artefacts -- such as stray light
spikes, unmasked blooming, prominent residual structures 
around bright but unsaturated stars -- cause deviations from the 
zero level. Such information is therefore useful for further filtering 
of outliers and associate quality flags to the photometric data points.
In addition to the aforementioned data, light curve files are extended with 
three additional columns showing the phase angle values, observer-centric 
distances and heliocentric distances.

\subsection{Residual spectra}
\label{sec:spectrum}

Residual spectra are generated by frequency scanning with a step size and
coverage in accordance with the TESS sector time-span and the TESS FFI
cadence, respectively. Namely, the total time-span of $\sim 27$ days on
average imply a stepsize of $\Delta f=0.01\,{\rm c/d}$ while the 
Nyquist criterion maximizes the scanning interval in 
$f_{\rm max}=24\,{\rm c/d}$. The residual spectrum is then computed for a
certain input frequency $f$ by minimizing the parameters
$A$, $B$, and $k_i$ ($i=0,1,2$) for the model function 
\begin{eqnarray}
m(t) & = &  \sum\limits_{i=0}^2 k_i[\alpha(t)-\alpha_0]^i + \\
& & + A\cos[2\pi f(t-T_0)]+B\sin[2\pi f(t-T_0)]. \nonumber
\end{eqnarray}
where $m(t)$ is the observed magnitude (corrected for the variations
in the solar and observer distances) at the instance $t$, $\alpha$ is the
phase angle, $\alpha_0$ is the mean phase angle throughout the observations,
and $T_0$ is an approximate mid-time of the observations. The actual 
values of $\alpha_0$ and $T_0$ do not alter the residuals (hence
the spectra), however, setting the aforementioned values helps to minimize
the numerical round-off errors and $k_0$ can also be interpreted as a mean
brightness magnitude throughout the observations. 

\subsection{Metadata}
\label{sec:metadata}

In the case of the light curve and residual spectrum analysis, metadata
represents the rotation frequency (and/or equivalently, the rotation period),
the characteristics of the light curve shape and the peak-to-peak 
amplitude as well as any associated external database. 
While the processing scripts store metadata in separate files in a form of key-value pairs, the final data product includes a list of 
concatenated metadata in a tabular form.

In addition, this metadata table is extended with various large 
asteroid database information for convenience and further analysis. This
information can be used to create additional types of statistics and 
have estimations for biases (see Sec.~\ref{sec:comparison} for examples).
In our published database, we included the most recent version
of the synthetic proper orbital elements of \cite{knezevic2000},
as available online\footnote{\url{https://newton.spacedys.com/astdys2/}},
the asteroid family catalog Version 3 of \cite{nesvorny2015} and
the most recent version of the Asteroid Lightcurve Database 
\citep[LCDB,][]{warner2009}. Of course, the overlap with neither of the
aforementioned databases are complete and there are only 1563 objects
for which both proper orbital elements and LCDB data are available.

\subsection{Validation plots}
\label{sec:validationplots}

For a quick manual vetting of the results of the photometric analysis,
we create a four-panel summary plot for each object. The four 
plots are the unfolded light curve, the residual spectrum, the 
folded light curve with the dominant period and the folded light curve with the double of the dominant period. 

\subsection{Object light curve plots and slides}
\label{sec:objectplots}

These plots contain the same information as the validation plots, but in a bit 
different arrangement and these display only a single folded light curve with the most likely rotation period. 
The plots also show this rotation period in the units of hours. 
We note here that the time instances for both the plots and all 
of the light curve data products are given in in Julian Days (JDs). 
As an example, two of such object light curve plots are displayed in 
Fig.~\ref{fig:objectplots} for the objects  (354) Eleonora 
and (220281) 2003\,BA$_{47}$. These objects represent the 
bright end and the faint end of our catalogue.

\section{Comparison with existing databases}
\label{sec:comparison}

\subsection{Asteroid Lightcurve Database -- LCDB}
\label{sec:lcdb}

The most comprehensive database available in the literature is the
Asteroid Lightcurve Database\footnote{\url{http://www.minorplanet.info/lightcurvedatabase.html}} \citep[LCDB, see][]{warner2009}. The most recent (August 2019) release of this database contains $4842$ objects for which a valid rotation
period {\it and} brightness variation amplitude is
associated\footnote{We note here that incomplete
amplitude information but settled rotation periods are available for
$20462$ objects.}. While this amount of data is nearly half of the 
entries available in the TESS minor planet data, the LCDB cites $2788$ 
bibliographic sources (concerning the entire database), therefore
one should consider the inhomogeneity while interpreting LCDB statistics. 
However, we expect that the aforementioned quality constraints of 
selecting $4842$ objects ensure the robustness of the data products.

\begin{figure*}[!ht]
\plottwo{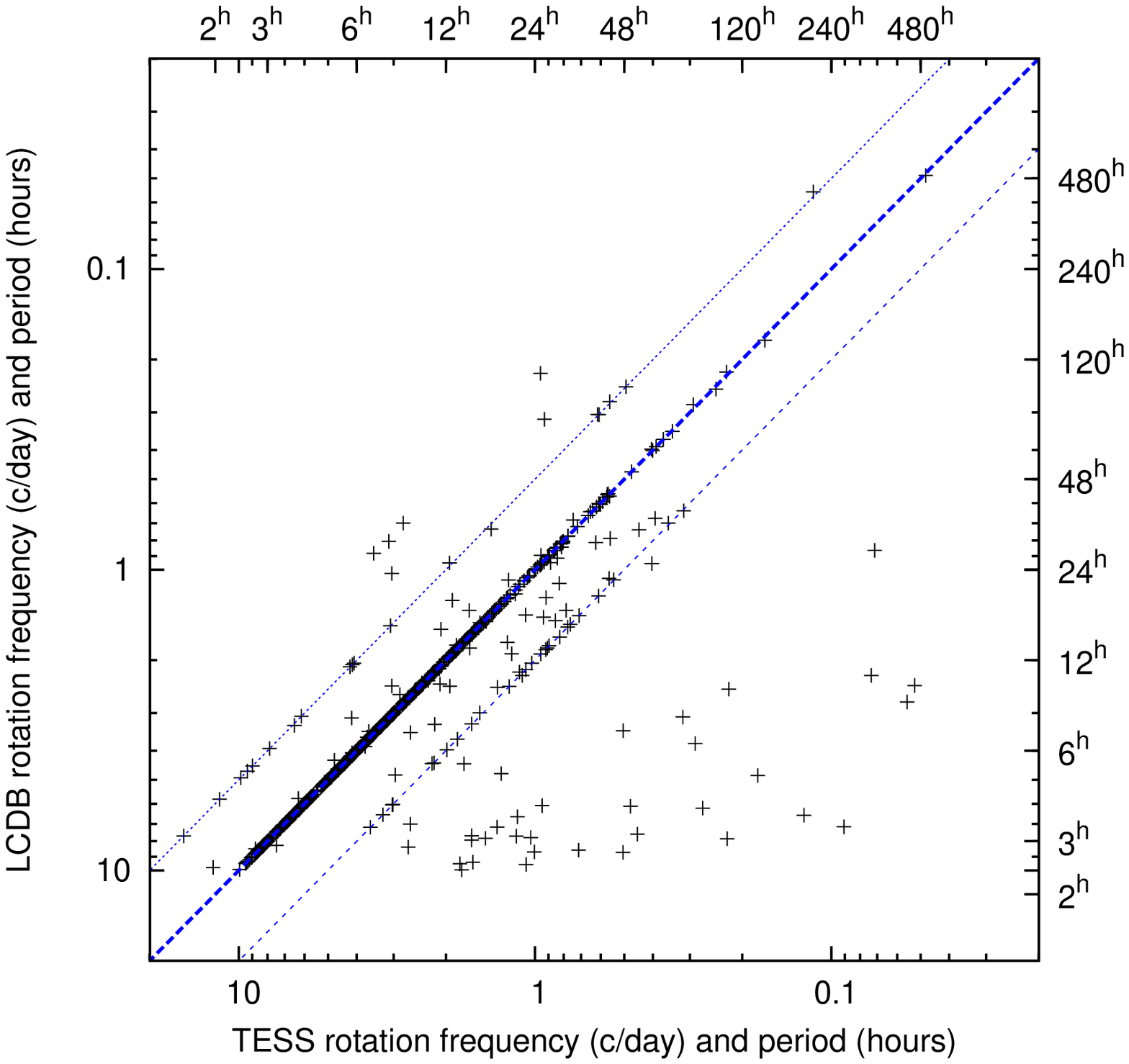}{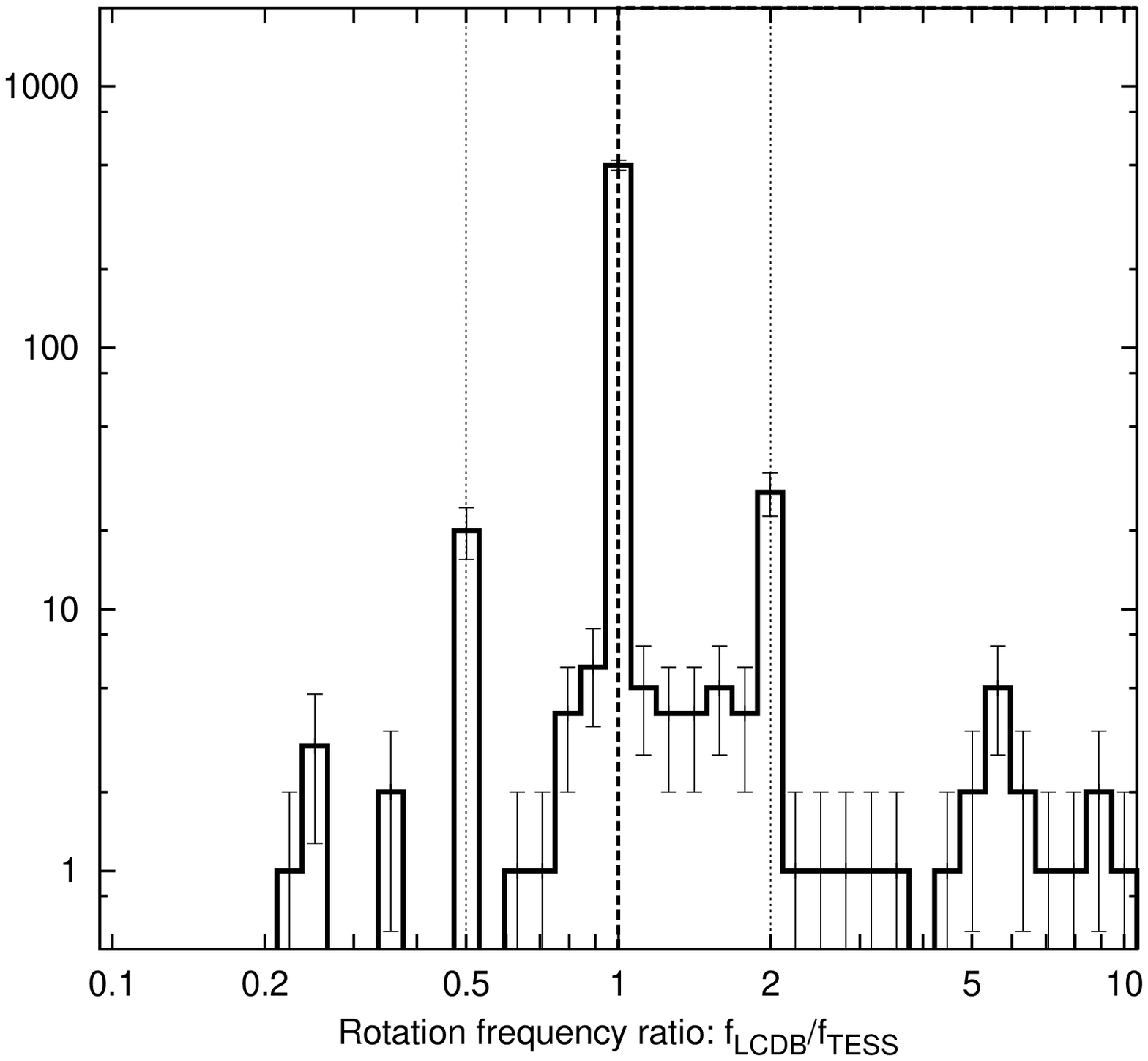}
\caption{The {\it left panel} shows the rotation periods of the 624 objects
for which reliable rotation characteristics (i.e. periods {\it and} 
amplitudes) are available both in the LCDB catalogue and the
TESS observations presented in this paper. The thick line and the two dashed
lines correspond to the same rotation frequencies as well as the $1:2$ and
$2:1$ ratios, respectively. The {\it right panel} displays the histogram 
of the frequency ratios of the objects available both in the LCDB 
catalogue and the presented TESS minor planet catalog. In total, $\sim80\%$
of the matched objects have the same derived rotation periods while
in the case of $\sim 8\%$ of the objects, the newly derived preferred 
periods are either the double or half of the periods available in the LCDB.
TESS measurements clearly identified longer rotation periods for 
the majority of the remaining $\sim 60$ objects.}
\label{fig:tesslcdb}
\end{figure*}

\begin{figure*}[!ht]
\plottwo{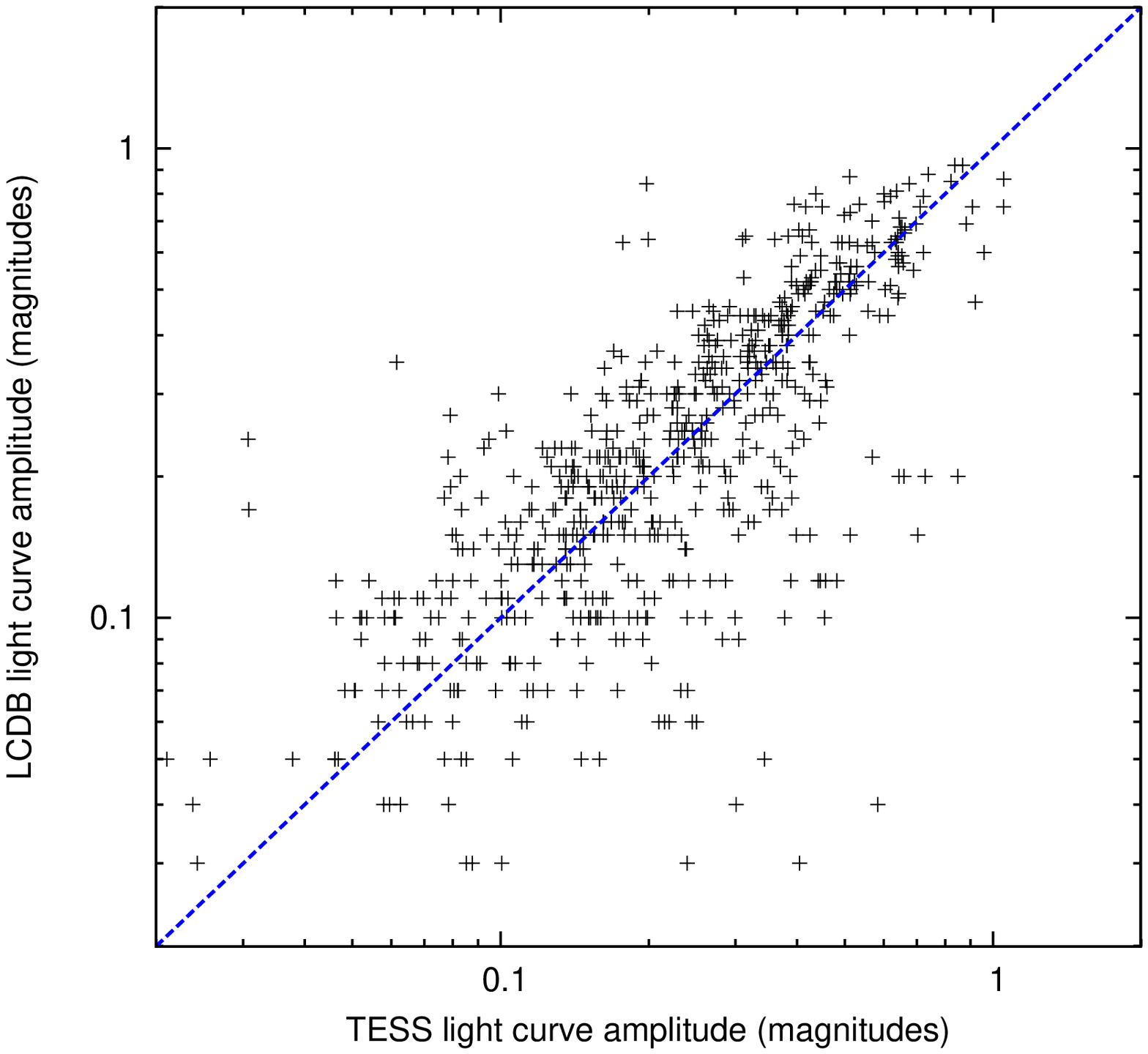}{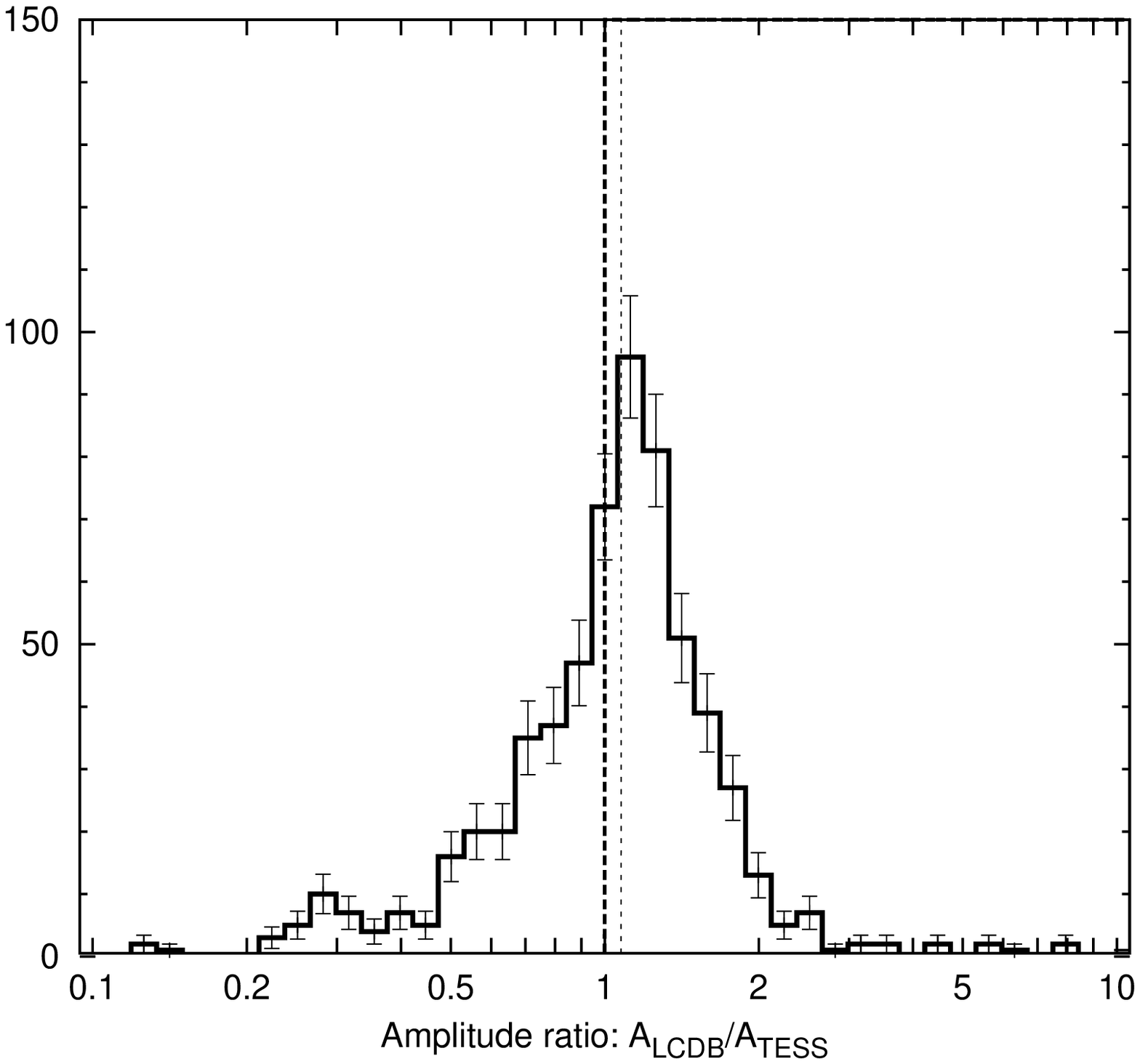}
\caption{{\it Left panel:} light curve peak-to-peak amplitudes 
for the 624 objects where rotation characteristics are available both 
in the LCDB catalogue and the TESS observations presented here. 
{\it Right panel:} the histogram of the distribution of the amplitude
ratios. The thick vertical line shows the unity ratio while the thin 
vertical dashed line at $\sim 1.076$ shows the median value of the 
amplitude ratios.}
\label{fig:tesslcdbamplitudes}
\end{figure*}

In total, we identified $624$ objects which are available both in
TSSYS-DR1 and LCDB (with sufficiently strong qualification). 
We note here that there are $1535$ objects available both in TSSYS-DR1 
and LCDB if we do not consider the amplitude quality criteria mentioned above.
In Figs.~\ref{fig:tesslcdb} and \ref{fig:tesslcdbamplitudes} we
displayed the rotation frequency and amplitude correlations, respectively, between the two databases. Considering the rotation periods, we found that the
agreement is perfect for $\sim80\%$ of the objects while there are
a few dozens of objects where the double-peaked ambiguity yields a
$1:2$ or $2:1$ ratio. The amount of such ambiguities is roughly the the
same (19 vs.\ 28) for the two ratios. Otherwise, it is worth to mention here
that TESS clearly identifies the objects with longer periods better, 
suspecting an unclear origin of the otherwise shorter reported
periodicity in LCDB (see the points above the $1:1$ and $2:1$ line on the
left panel of Fig.~\ref{fig:tesslcdb} or the histogram distribution
at the right tail on the right panel of the same Figure). 

Regarding to the interpretation of the correlations between amplitudes (see 
Fig.~\ref{fig:tesslcdbamplitudes}), the larger amplitudes present in the LCDB 
is a clear signature of the bias in the TESS observations. Namely, TESS 
observes minor planets close to the opposition, i.e. at small phase angles 
while LCDB contains many kinds of observations (yielding better coverage in phase angles), not just ones close to the opposition. According to the
expectations \citep{zappala1990}, higher phase angles would yield higher
amplitudes, which can explain the shift in the correlation diagram and the 
corresponding histogram. However, one should note that because of this 
TESS-specific observing constraint as well as due to the fact that the 
presented data release contains only a single epoch while LCDB aggregates 
data from many observing runs, such a statistical comparison between TESS 
and LCDB amplitudes needs to be considered tentative. While the presented 
TESS data series are highly homogeneous, it shows an amplitude characteristics 
only for a single observing geometry, leaving many aspects 
of shape characteristics ambiguous.

\begin{figure}[!ht]
\plotone{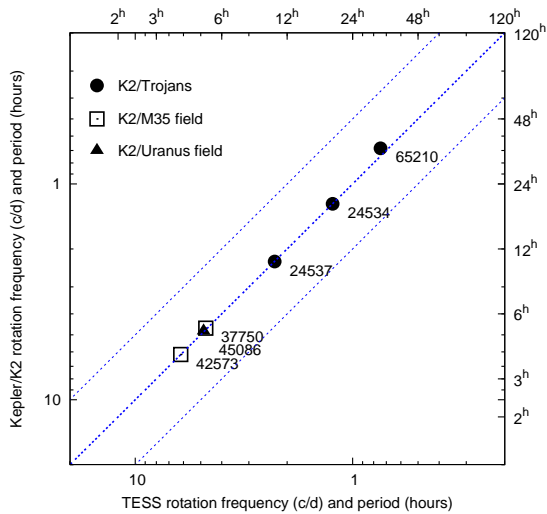}
\caption{The 6 minor planets for which both Kepler/K2 and TESS measurements
are available. 3 out of these 6 objects are Jupiter Trojans while the another
two are main-belt asteroids. With the exception of the Trojan asteroid
(65210) Stichius, the periods match within $1.5\%$. The agreement 
for (24534) 2001\,CX$_{27}$ and (42573) 1997\,AN$_{1}$ are less than
one tenth of a percent. (65210) Stichius show a difference of $\sim 8\%$
between the derived periods.}
\label{fig:k2tess}
\end{figure}

\subsection{K2 Solar System Studies -- K2SSS}
\label{sec:k2sss}

While having scanned various fields close to the ecliptic plane,
the K2 mission \citep{howell2014} also provided a highly efficient way to
provide uninterrupted observations for various classes of Solar System objects.
These classes include not only main-belt and Trojan asteroids but 
trans-Neptunian objects \citep{pal2015}, irregular satellites of 
giant planets \citep{kiss2016,farkas2017}, and the Pluto-Charon system
\citep{benecchi2018}. K2 observations also implied the discovery of the 
satellite of (225088) 2007\,OR$_{10}$ \citep{kiss2017} when its
slow rotation was detected \citep{pal2016}.

With the exception of the discovery and 
photometry of the trans-Neptunian object (506121) 2016\,BP$_{81}$
\citep{barensten2017}, all of these object classes were measured as targeted
observations, i.e. with pre-allocated K2 target pixel files (arranged into 
special boomerang-shaped pixel blocks). In the case of main-belt and Trojan
asteroids, there are examples of targeted 
observations \citep{marciniak2019,szabo2017,ryan2017}
as well as photometry on contiguous superstamps \citep{szabo2016,molnar2018}
when asteroids serendipitously crossed these celestial areas. However,
the data reduction pattern does not differ significantly for 
pre-allocated reductions and the analysis of contiguous superstamps 
with the exception of the aforementioned querying of the objects (by
tools like \texttt{EPHEMD}) in the latter case. See, e.g., 
\cite{szabo2017} for a detailed description about the data reduction
for K2 minor planet observations observations. 

In order to compare the objects observed by any initiative 
of the K2 Solar System Surveys with this recent TESS-based photometry,
we identifies 6 main-belt and Trojan objects that were observed both by
K2 and TESS. These were      
(24534) 2001\,CX$_{27}$,
(24537) 2001\,CB$_{35}$,
(37750) 1997\,BZ,
(42573) 1997\,AN$_{1}$,
(45086) 1999\,XE$_{46}$ and
(65210) Stichius. We found that the derived rotation periods match
perfectly in $5$ of the $6$ cases, see Fig.~\ref{fig:k2tess}. There
was only a slight offset for (65210) Stichius, due to its faintness and
long rotation period of $\sim 32$\, hours. 

\begin{figure}[!ht]
\plotone{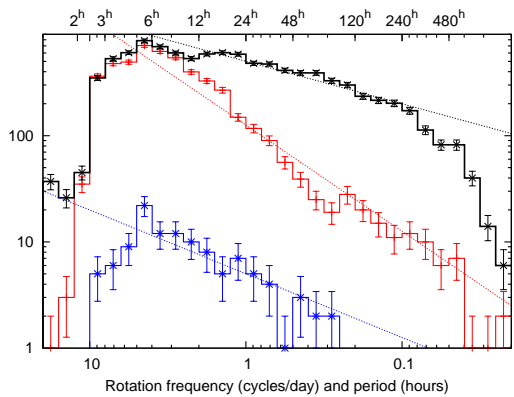}
\caption{The number of objects as the function of their periods,
provided by various databases. The black curve shows the
period distribution for TSSYS-DR1 ($9912$ objects), the red curve 
shows the period distribution for the $4842$ LCDB objects for 
which a valid rotation period and brightness variation 
amplitude have been derived at the same time. The blue curve shows 
the period distribution for $113$ serendipitous 
main-belt asteroids provided by the analysis of three 
K2 superstamps. The thin dashed lines guide the eye to provide
a tentative slope at the long period (low frequency) parts
of these distributions.}
\label{fig:periods}
\end{figure}

\subsection{Period statistics}
\label{sec:periods}

In Fig.~\ref{fig:periods} we displayed the histograms of the detected rotation 
periods for this TESS-based asteroid survey, the LCDB and the K2 serendipitous 
main-belt asteroid detections on the M35 and Neptune-Nereid 
fields \citep{szabo2016}, as well as on the Uranus field \citep{molnar2018}. A tentative
fit in the long-period part of these histograms clearly show that both ground-based and
shorter duration but otherwise uninterrupted space-borne measurements underestimate
the number of objects in the population of slow rotators. Therefore, we can safely 
conclude that the nearly one-month long continuous data acquisition of
TESS would provide us the most unbiased coverage and confirmation 
of slowly rotating asteroids. However, it is still an interesting question where the cut-off of TESS is, above which the rotation period statistics become significantly biased. 
The divergence between the LCDB and TESS histograms stars at rotation periods of $8-10$\,hours. Below this period, the two statistics nicely agree down to the periods
of $\sim 2$\,hours range.

\section{Summary}
\label{sec:summary}

In this paper we presented the first data release of the complete Southern Survey of the Transiting Exoplanet Survey Satellite in terms of analysis of bright, main-belt and Trojan asteroids crossing the field-of-view of Camera \#1. This survey triples the number of asteroids with accurately determined rotation characteristics. Another advantage of the presented catalogue is that it is fully homogeneous considering both data acquisition and data processing principles.  Further fine-tuning in the pipeline presented here is also possible, and we have the intention to process and add further object classes, including 
Centaurs, trans-Neptunian objects and near-Earth objects \citep[see also][]{milam2019}.

TESS is now observing the Northern Hemisphere, opening the possibilities to re-observe many of the objects
presented in this data release with a completely different observing geometry with respect to the spin-axis 
orientation of these bodies. Such further observations would help us to interpret the derived light curve 
characteristics, specifically the amplitude in a more accurate manner and therefore helping the 
analysis for a more accurate comparison with LCDB.} We should also express our hope that 
the extended mission of TESS would include wide coverage of the ecliptic plane, further 
expanding our collection of asteroid observations and increase the number of multi-epoch observations.  

\facilities{TESS \citep{ricker2015}, \textit{Gaia} DR2 \citep{gaia2018}} \software{FITSH \citep{pal2012}, EPHEMD}

\vspace*{2mm}

\begin{acknowledgements}
We would like to thank Brian D. Warner for the careful review of our paper and his highlights of many aspects of light curve interpretation and caveats. A.P.\ would like to thank Matt Holman, George Ricker, Roland Vanderspek, Joel Villasenor and Deborah Woods for the fruitful discussions about the astrometry of TESS full-frame images and Solar System topics in general. This paper includes data collected by the TESS mission. Funding for the TESS mission is provided by the NASA Explorer Program. This project has been supported by the Lend\"ulet Program  of the Hungarian Academy of Sciences, project No. LP2018-7/2019. Additional support is received from the K-125015 and GINOP~2.3.2-15-2016-00003 grants of the National Research, Development and Innovation Office (NKFIH, Hungary). Zs.B.\ acknowledges the support provided from the National Research, Development and Innovation Fund of Hungary, financed under the PD$_{17}$ funding scheme, project no. PD-123910. L.M.\ was supported by the Premium Postdoctoral Research Program of the Hungarian Academy of Sciences. Cs.K.\ was supported by the \'UNKP-19-2 New National Excellence Program of the Ministry of Human Capacities. Partial funding of the computational infrastructure and database servers are received from the grant KEP-7/2018 of the Hungarian Academy of Sciences. Gy.M.Sz. was supported by the Hungarian NKFI Grant  K-119517 and the City of Szombathely under Agreement No. 67.177-21/2016. The work of G.M. was supported by the PD-128360 project of the National Research, Development and Innovation Office, Hungary. This work has made use of data from the European Space Agency (ESA) mission {\it Gaia}, processed by the {\it Gaia} Data Processing and Analysis Consortium (DPAC). Funding for the DPAC has been provided by national institutions, in particular the institutions participating in the {\it Gaia} Multilateral Agreement.
\end{acknowledgements}

%\bibliographystyle{aj}
%\bibliography{K2asteroids}

{}

\end{document}